\documentclass[a4paper,11pt]{article}

\pdfoutput=1
\usepackage{jheppub}

\usepackage{graphicx}
\usepackage{subfig}

\usepackage{amsmath}
\usepackage{xspace}
\usepackage{xcolor}
\usepackage{float}
\usepackage{bbold}

\def\nn{\nonumber \\ }
\usepackage{enumerate}

\newcommand{\df}{\mathrm{d}}

\newcommand{\eq}[1]{Eq.~\eqref{eq:#1}}

\renewcommand{\sec}[1]{Sec.~\ref{sec:#1}}

\newcommand{\subsec}[1]{Sec.~\ref{subsec:#1}}

\newcommand{\fig}[1]{Fig.~\ref{fig:#1}}

\title{Resummation of electroweak Sudakov logarithms for real radiation}

\author{Christian W Bauer, }
\author{Nicolas Ferland}
\affiliation{Ernest Orlando Lawrence Berkeley National Laboratory, University of California, Berkeley, CA 94720, USA}
\emailAdd{cwbauer@lbl.gov}
\emailAdd{nferland@lbl.gov}

\abstract{
Using the known resummation of virtual corrections together with knowledge of the leading-log structure of real radiation in a parton shower, we derive analytic expressions for the resummed real radiation after they have been integrated over all of phase space. Performing a numerical analysis for both the 13 TeV LHC and a 100 TeV $pp$ collider, we show that resummation of the real corrections is at least as important as resummation of the virtual corrections, and that this resummation has a sizable effect for partonic center of mass energies exceeding $\sqrt{s} = {\cal O}$(few TeV). For partonic center of mass energies $\sqrt{s} \gtrsim$ 10 TeV, which can be reached at a 100 TeV collider, resummation becomes an O(1) effect and needs to be included even for rough estimates of the cross-sections. 
}
\vspace{10cm}

\begin{document}

\maketitle

\section{Introduction}\label{sec:intro}
It is well known that perturbative corrections in scattering processes have infrared (IR) sensitivity to soft and collinear emissions. For electroweak corrections involving massive $Z$ and $W$ bosons, the mass of the vector boson regulates the IR divergences, such that the IR sensitivity yields logarithmic dependence on the ratio of the vector boson mass over the partonic center of mass energy
\begin{align}
L_V \equiv \ln m_V^2 / s
\,.
\end{align}
Both virtual and real corrections give rise to such logarithmic corrections, and  as usual, their effect cancels in fully inclusive observables. 

This situation is very similar to the more common situation of gauge theories with massless gauge bosons, such as QCD, for which both virtual and real corrections are IR divergent, and the divergence cancels in the sum that enters in any physical observable. However, there are two main differences: First, for massive gauge bosons, both virtual and real contribution are separately finite (albeit with logarithmic dependence on the ratio of the mass of the vector boson to the center of mass energy), such that there is no need to combine them for physical observables. Second, even if the measurement is completely inclusive over the final state, the initial beams of colliders are typically not SU(2) singlets, such that one can never perform a fully inclusive measurement. Thus, essentially any measurement has logarithmic sensitivity to the ratio $m_V^2 / s$, such that for high enough center of mass energies electroweak corrections become very large. 

Much effort has been put into understanding the electroweak logarithms arising from virtual corrections~\cite{Kuhn:1999nn,Fadin:1999bq,Ciafaloni:1999ub,Beccaria:2000jz,Hori:2000tm,Ciafaloni:2000df,Denner:2000jv,Denner:2001gw,Melles:2001ye,Beenakker:2001kf,Denner:2003wi,Pozzorini:2004rm,Feucht:2004rp,Jantzen:2005xi,Jantzen:2005az,Jantzen:2006jv,Chiu:2007yn,Chiu:2008vv,Manohar:2012rs}. A general result for the logarithmic dependence at one-loop has been derived in~\cite{Denner:2000jv,Denner:2001gw}, and a general method for the resummation of these logarithms has been developed in~\cite{Chiu:2007yn,Chiu:2008vv}. This method uses SCET which has been developed in~\cite{Bauer:2001ct,Bauer:2000ew,Bauer:2000yr,Bauer:2001yt}. While real radiation of electroweak gauge bosons typically leads to different experimental signatures, in many cases some amount of real radiation is included in the event samples. For example, in many analyses missing energy is not vetoed, such that radiation of $Z$ bosons which decay to neutrinos is included, and any analysis that is inclusive over the number of jets includes real radiation of hadronically decaying vector bosons. 

Logarithmic dependence from real emission, however, has been studied much less. The ${\cal O}(\alpha)$ correction from real emission can of course easily be included using phase space integrations over tree-level matrix elements, and a study of the effect of such real radiation on many experimental observables was performed in~\cite{Baur:2006sn}. It was found that the real corrections can have a sizable effect, albeit typically not quite as large as the virtual corrections. The effects of real radiation have  recently been studied in~\cite{Manohar:2014vxa,Bell:2010gi}.

In this paper we consider the Drell-Yan cross sections
$\sigma_{p p \to \ell_1 \ell_2}$  and $\sigma_{p p \to \ell_1 \ell_2 V} \,.$
Here $\ell$ denotes either an electron, positron or a neutrino, and $V$ denotes either a $Z$, a photon or a $W$ boson. For Drell-Yan production at NLO accuracy, the first cross section includes the tree-level and the one-loop virtual correction, while the second cross section denotes the real radiation of a gauge boson. Collectively, we will represent these two cross sections as 
$\sigma_{p p \to \ell_1 \ell_2 X}\,.$

To calculate this hadronic scattering cross-sections one starts as always by factorizing the short distance partonic scattering from the non-perturbative physics describing the binding of partons into hadrons
\begin{align}
\sigma_{p p \to {\rm final}} &= \sum_{ab} \int \!\df x_a\df x_b \, f_{a/p}(x_a) f_{b/p}(x_b)
\,  \hat \sigma_{ab \to {\rm final}}(x_a,x_b)
\,,\end{align}
where $f_{i/p}(x_i)$ are the parton distribution functions to find parton $i$ in the proton $p$ with momentum fraction $x_i$, and $\hat \sigma_{ab \to {\rm final}}(x_a,x_b)$ denotes the partonic scattering cross-section. By Lorentz invariance, the dependence of the partonic cross-sections on the momentum fractions $x_i$ is only through the product $s = x_a x_b S$, and not through the rapidity $Y = \ln (x_a / x_b) / 2$. Thus, one can write 
\begin{align}
\label{eq:hadronicDY}
\frac{\df \sigma_{p p \to {\rm final}} }{\df s} = \sum_{ab} {\cal L}_{ab} (s) \, \hat \sigma_{ab \to {\rm final}}(s)
\,,\end{align}
where the parton luminosity is defined as
\begin{align}
{\cal L}_{ab} (s) =  \int \!\df x_a\df x_b \, f_{a/p}(x_a) f_{b/p}(x_b) \, \delta(s - x_a x_b S)
\,.\end{align}

It is the purpose of this paper to derive simple analytical expressions for the resummation of electroweak Sudakov logaritms at leading logarithmic (LL) accuracy. To achieve this, we combine knowledge of the resummation of the virtual corrections, together with the known IR structure of real radiation from parton showers. This allows us to obtain simple analytical results which we present here. 
This approach is similar to the one used in \cite{Fadin:1999bq}, where these effects were discussed focussing on $e^+ e^-$ collider.
We then use these results to give expressions for the individual contributions to the hadronic cross-sections $\sigma_{p p \to \ell_1 \ell_2 X}$ as a function of the center of mass energy $\sqrt{s}$. We reproduce the well known fact that resummation has a large effect on the virtual contribution for $\sqrt{s} \gtrsim 2$ TeV. The resummation of the real corrections changes the fixed order results by more than 40\% for  $\sqrt{s} \gtrsim 2$ TeV, and this effect grows to about  200\% for $\sqrt{s} \sim 25$ TeV. 

This paper is organized as follows: In \sec{structureOfLogs} we set up some notation and derive general expressions for the cross-sections which highlight the structure of the logarithmic corrections. In \sec{FO} we give the (well known) fixed order results of the virtual and real electroweak corrections which are enhanced by two powers of the logarithms $L_V$.  In \subsec{VirtLL} we give the resummed results for the virtual corrections, keeping only LL accuracy. In \subsec{RealInclLL}, we derive the resummation of the leading logarithmic terms for real corrections. In \sec{Results} we provide a numerical analyses of these results, both for the 13 TeV LHC and a 100 TeV $pp$ collider. We finish with Conclusions in \sec{Conclusions}. 
\section{The structure of logarithmic terms in the perturbative expansion}\label{sec:structureOfLogs}
To first order in electroweak perturbation theory the partonic scattering cross-section can be written as
\begin{align}
\label{eq:RealVirtSplit}
\hat \sigma_{ij \to \ell_1 \ell_2 X} = \hat \sigma^{(0)}_{ij \to \ell_1 \ell_2} +  \hat \sigma^{(1)}_{ij \to \ell_1 \ell_2 X}
\,,\end{align}
where $\hat \sigma^{(0)}_{ij \to \ell_1 \ell_2}$ denotes the Born cross-section, and $\hat \sigma^{(1)}_{ij \to \ell_1 \ell_2 X}$ the ${\cal O}(\alpha)$ correction. This first order perturbative correction can be decomposed into a virtual and real contribution, and each of these two terms can be further separated by the flavor of the vector boson in the loop or real final state. This gives
\begin{align}
\label{eq:RealVirtSplit}
\hat \sigma^{(1)}_{ij \to \ell_1 \ell_2 X} = \sum_V \left[\hat \sigma^{V}_{ij \to \ell_1 \ell_2}  + \hat \sigma_{ij \to \ell_1 \ell_2 V}\right]
\,.
\end{align}
Here the first term describes the contribution from one-loop diagrams with $V = \{Z, W^\pm, \gamma\}$ in the loop, while the second term is given by the real radiation of an electroweak gauge boson $V$. 

Each of the terms on the right hand side of \eq{RealVirtSplit}, contains double and single logarithmic terms of the form $\alpha \, L_V$ and $\alpha \, L_V^2$. By the KLN theorem \cite{Kinoshita:1962ur}, these logarithmically enhanced terms cancel when we sum over complete gauge multiplets. Thus, the completely inclusive cross-section has no logarithmically enhanced terms (called ``finite'' below)
\begin{align}
\label{eq:AllCancellation}
\sum_{a,b,\ell_1,\ell_2} \hat \sigma^{(1)}_{ab \to \ell_1 \ell_2 X} = {\rm finite}
\,.\end{align}
The sum of the initial state is over $a,b \in \{u, d\}$,  that of the final state is over $\ell_1,\ell_2 \in \{e^+, e^-, \nu\}$, and we sum over the virtual contributions, and well as the real contributions with either a $Z$, a photon or $W$ in the final state. 
Since the logarithmic enhancement depends on the mass of the vector boson, the logarithmic corrections from $Z$ bosons, photons, and $W$ bosons cancel separately between real and virtual corrections. Furthermore, since the emission of a $Z$ boson or a photon does not change the flavor of a fermion, the cancellation of the terms enhanced by $L_Z$ happens for each flavor assignment separately. This gives
\begin{align}
\label{eq:ZCancellation}
\hat \sigma^{Z}_{ab \to \ell_1 \ell_2}  + \hat \sigma_{ab \to \ell_1 \ell_2 Z} = {\rm finite} \,, \qquad 
\hat \sigma^{\gamma}_{ab \to \ell_1 \ell_2}  + \hat \sigma_{ab \to \ell_1 \ell_2 \gamma} = {\rm finite}
\,.
\end{align}
The emission of $W$ bosons, on the other hand, changes the flavor of the fermion. This implies that the cancellation of the terms enhance by $L_W$ only happens after summing over all flavors
\begin{align}
\label{eq:WCancellation}
\sum_{a,b,\ell_1,\ell_2} \left[ \hat \sigma^{W}_{ab \to \ell_1 \ell_2}  + \hat \sigma_{ab \to \ell_1 \ell_2 W}\right] = {\rm finite}
\,.
\end{align}

An important consequence of~\eq{WCancellation} is that a finite answer is only obtained if one sums over all initial state flavors. Thus, the hadronic cross-section given in \eq{hadronicDY} still contains contributions enhanced logarithmically by factors of $L_W$. This is because each initial state flavor is multiplied by a different parton luminosity, such that the sum over initial state flavors cannot be observed.

\section{Fixed order calculations of the virtual and real contributions}\label{sec:FO}
In this section we consider the fixed order expansion of the virtual corrections $\hat \sigma^{V}_{ab \to \ell_1 \ell_2}$ and the real contributions $\hat \sigma_{ab \to \ell_1 \ell_2 V}$. For both terms, we quote here only the dominant contributions that are enhanced by two powers of $L_V$.

For completeness, we start with the Born cross-sections, including the contributions from both photon and $Z$ exchange for the neutral initial and final states, and from $W$ exchange in the charged cases. They  are given by
\begin{align}
\label{eq:Born}
\hat \sigma^B_{u\bar u \to e^-e^+} &= N \,  \frac{85 \alpha_1^2 + 6 \alpha_1 \alpha_2 + 9 \alpha_2^2}{54}
\nn
\hat \sigma^B_{u\bar u \to \nu\bar\nu} &= N \,  \frac{17 \alpha_1^2 - 6 \alpha_1 \alpha_2 + 9 \alpha_2^2}{54}
\nn
\hat \sigma^B_{d\bar d \to e^-e^+} &= N \,  \frac{25 \alpha_1^2 - 6 \alpha_1 \alpha_2 + 9 \alpha_2^2}{54}
\nn
\hat \sigma^B_{d\bar d \to \nu \bar\nu} &= N \,  \frac{5 \alpha_1^2 + 6 \alpha_1 \alpha_2 + 9 \alpha_2^2}{54}
\nn
\hat \sigma^B_{u\bar d \to \nu e^+} &= N \,  \frac{2\alpha_2^2}{3}
\nn
\hat \sigma^B_{d\bar u \to e^-\nu} &= N \,  \frac{2\alpha_2^2}{3}
\,,\end{align}
where we have defined the constant
\begin{align}
N = \frac{\pi}{8 N_C s}
\,,
\end{align}
with $N_C = 3$ denoting the number of colors. To simplify the notation, we have written our results in terms of the coupling constants of the unbroken theory
\begin{align}
\alpha_1 = \frac{\alpha_{\rm em}}{\cos^2\theta_W} \,, \qquad \alpha_2 = \frac{\alpha_{\rm em}}{\sin^2\theta_W}
\,,\end{align}
where $\theta_W$ is the weak mixing angle and $\alpha_{\rm em}$ is the electromagnetic coupling constant, also known as the fine-structure constant. All coupling constant are evaluated at the scale $\sqrt{s}$, the center of mass energy, $\alpha_i=\alpha_i(\sqrt{s})$.

The virtual corrections from $W$ exchange that are enhanced by two powers of the logarithm are easily obtained for example using \cite{Denner:2000jv} and are given by
\begin{align}
\label{eq:FOVirtW}
\hat \sigma^{W^\pm}_{u\bar u \to e^-e^+} &= - N \,  \frac{\alpha_W \, L_W^2}{4\pi} \frac{11 \alpha_1^2 + 6 \alpha_1 \alpha_2 + 9 \alpha_2^2}{27}
\nn
\hat \sigma^{W^\pm}_{u\bar u \to \nu\bar\nu} &= - N \,  \frac{\alpha_W \, L_W^2}{4\pi} \frac{3 \alpha_1^2 -2 \alpha_1 \alpha_2 + 3 \alpha_2^2}{9}
\nn
\hat \sigma^{W^\pm}_{d\bar d \to e^-e^+} &= - N \,   \frac{\alpha_W \, L_W^2}{4\pi} \frac{5 \alpha_1^2 - 6 \alpha_1 \alpha_2 + 9 \alpha_2^2}{27}
\nn
\hat \sigma^{W^\pm}_{d\bar d \to \nu \bar\nu} &= - N \,   \frac{\alpha_W \, L_W^2}{4\pi} \frac{\alpha_1^2 + 2 \alpha_1 \alpha_2 + 3 \alpha_2^2}{9}
\nn
\hat \sigma^{W^\pm}_{u\bar d \to \nu e^+} &= - N \,  \frac{\alpha_W \, L_W^2}{4\pi} \frac{4 \alpha_2^2}{3}
\nn
\hat \sigma^{W^\pm}_{d\bar u \to e^-\nu} &= - N \,  \frac{\alpha_W \, L_W^2}{4\pi} \frac{4 \alpha_2^2}{3}
\,.\end{align}
Those from $Z$ exchange are given by
\begin{align}
\label{eq:FOVirtZ}
\hat \sigma^Z_{u\bar u \to e^-e^+} &= - N \,  \frac{\alpha_Z \, L_Z^2}{4\pi} \frac{(99-366s_W^2+2210s_W^4)\alpha_1^2 + (9 - 30s_W^2+26s_W^4) \left(6 \alpha_1 \alpha_2 + 9\alpha_2^2 \right)}{486}
\nn
\hat \sigma^Z_{u\bar u \to \nu\bar\nu} &= - N \,   \frac{\alpha_Z \, L_Z^2}{4\pi} 
\frac{(81-12s_W^2+136s_W^4)\alpha_1^2 + (9 - 12s_W^2+8s_W^4) \left(-6 \alpha_1 \alpha_2 + 9\alpha_2^2 \right)}{486}
\nn
\hat \sigma^Z_{d\bar d \to e^-e^+} &= - N \,    \frac{\alpha_Z \, L_Z^2}{4\pi} 
\frac{5(9-24s_W^2+100s_W^4)\alpha_1^2 + (9 - 24s_W^2+20s_W^4) \left(-6 \alpha_1 \alpha_2 + 9\alpha_2^2 \right)}{486}
\nn
\hat \sigma^Z_{d\bar d \to \nu \bar\nu} &= - N \,    \frac{\alpha_Z \, L_Z^2}{4\pi} 
\frac{(27-6s_W^2+10s_W^4)\alpha_1^2 + (9 - 6s_W^2+2s_W^4) \left(6 \alpha_1 \alpha_2 + 9\alpha_2^2 \right)}{486}
\nn
\hat \sigma^Z_{u\bar d \to \nu e^+} &= - N \,  \frac{\alpha_Z \, L_Z^2}{4\pi} \frac{2(9-18s_W^2+14s_W^4) \alpha_2^2}{27}
\nn
\hat \sigma^Z_{d\bar u \to e^-\nu} &= - N \,  \frac{\alpha_Z \, L_Z^2}{4\pi} \frac{2(9-18s_W^2+14s_W^4) \alpha_2^2}{27}
\,.
\end{align}
The virtual corrections from photon exchange that are enhanced by two powers of the logarithm are easily obtained and are given by
\begin{align}
\label{eq:FOVirtPhoton}
\hat \sigma^\gamma_{u\bar u \to e^-e^+} &= - N \,  \frac{\alpha_{em} \, L_\gamma^2}{4\pi} \frac{85 \alpha_1^2 + 6 \alpha_1 \alpha_2 + 9 \alpha_2^2}{54} \frac{26}{9}
\nn
\hat \sigma^\gamma_{u\bar u \to \nu\bar\nu} &= - N \,  \frac{\alpha_{em} \, L_\gamma^2}{4\pi} \frac{17 \alpha_1^2 -6 \alpha_1 \alpha_2 + 9 \alpha_2^2}{54} \frac{8}{9}
\nn
\hat \sigma^\gamma_{d\bar d \to e^-e^+} &= - N \,  \frac{\alpha_{em} \, L_\gamma^2}{4\pi} \frac{25 \alpha_1^2 -6 \alpha_1 \alpha_2 + 9 \alpha_2^2}{54} \frac{20}{9}
\nn
\hat \sigma^\gamma_{d\bar d \to \nu \bar\nu} &= - N \,  \frac{\alpha_{em} \, L_\gamma^2}{4\pi} \frac{5 \alpha_1^2 -6 \alpha_1 \alpha_2 + 9 \alpha_2^2}{54} \frac{2}{9}
\nn
\hat \sigma^\gamma_{u\bar d \to \nu e^+} &= - N \,  \frac{\alpha_{em} \, L_\gamma^2}{4\pi} \frac{2 \alpha_2^2}{3} \frac{14}{9}
\nn
\hat \sigma^\gamma_{d\bar u \to e^-\nu} &= - N \,  \frac{\alpha_{em} \, L_\gamma^2}{4\pi} \frac{2 \alpha_2^2}{3} \frac{14}{9}
\,,\end{align}
where the logarithms depend on a scale $\Lambda$, below which a photon is no longer resolved,
\begin{align}
L_\gamma \equiv \ln \left( \frac{\Lambda^2}{s} \right)
\,.\end{align}
We have also defined $s_W = \sin(\theta_W)$ and 
\begin{align}
\alpha_W = \alpha_2\,,  \qquad  \alpha_Z = \frac{\alpha_2}{\cos^2\theta_W}
\,.
\end{align}

The double-logarithmically enhanced real corrections with a $Z$ boson or a photon in the final state are related to the corresponding virtual corrections via
\begin{align}
\label{eq:FORealZ}
\hat \sigma_{q_1\bar q_2 \to \ell_1\ell_2 Z} = - \hat \sigma^Z_{q_1\bar q_2 \to \ell_1\ell_2}
\,,\end{align}
\begin{align}
\label{eq:RealPhoton}
\hat \sigma_{q_1\bar q_2 \to \ell_1\ell_2 \gamma} = - \hat \sigma^\gamma_{q_1\bar q_2 \to \ell_1\ell_2}
\,,\end{align}
which directly follow from \eq{ZCancellation},
for those with a $W^+$ boson in the final state we find
\begin{align}
\label{eq:RealWp}
\hat \sigma_{u \bar d \to e^- e^+ W^+} &= N \,  \frac{\alpha_W \, L_W^2}{4\pi} 
\frac{5\alpha_1^2+27\alpha_2^2}{54}
\nn
\hat \sigma_{u \bar d \to \nu \bar \nu W^+} &= N \,  \frac{\alpha_W \, L_W^2}{4\pi} 
\frac{\alpha_1^2+27\alpha_2^2}{54}
\nn
\hat \sigma_{u \bar u \to e^- \bar \nu W^+} &= N \,  \frac{\alpha_W \, L_W^2}{4\pi} 
\frac{17\alpha_1^2+27\alpha_2^2}{54}
\nn
\hat \sigma_{d \bar d \to e^- \bar \nu W^+} &= N \,  \frac{\alpha_W \, L_W^2}{4\pi} 
\frac{5\alpha_1^2+27\alpha_2^2}{54}
\,,
\end{align}
while those with a $W^-$ are given by
\begin{align}
\label{eq:RealWm}
\hat \sigma_{d \bar u \to e^- e^+ W^-} &= N \,  \frac{\alpha_W \, L_W^2}{4\pi} 
\frac{5\alpha_1^2+27\alpha_2^2}{54}
\nn
\hat \sigma_{d \bar u \to \nu \bar \nu W^-} &= N \,  \frac{\alpha_W \, L_W^2}{4\pi} 
\frac{\alpha_1^2+27\alpha_2^2}{54}
\nn
\hat \sigma_{u \bar u \to \nu e^+ W^-} &= N \,  \frac{\alpha_W \, L_W^2}{4\pi} 
\frac{17\alpha_1^2+27\alpha_2^2}{54}
\nn
\hat \sigma_{d \bar d \to \nu e^+ W^-} &= N \,  \frac{\alpha_W \, L_W^2}{4\pi} 
\frac{5\alpha_1^2+27\alpha_2^2}{54}
\,.\end{align}

From these results one can obtain the inclusive cross-sections, which have been summed over all final state flavors
\begin{align}
\label{eq:InclusivePartonic}
\hat \sigma_{u\bar u} = - N \,  \frac{\alpha_W \, L_W^2}{4\pi} \frac{\alpha_1^2-3\alpha_2^2}{9}
\nn
\hat \sigma_{d\bar d} = - N \,  \frac{\alpha_W \, L_W^2}{4\pi} \frac{\alpha_1^2-3\alpha_2^2}{9}
\nn
\hat \sigma_{u\bar d} = N \,  \frac{\alpha_W \, L_W^2}{4\pi} \frac{\alpha_1^2-3\alpha_2^2}{9}
\nn
\hat \sigma_{d\bar u} = N \,  \frac{\alpha_W \, L_W^2}{4\pi} \frac{\alpha_1^2-3\alpha_2^2}{9}
\,.\end{align}
From these expressions one can very easily validate that in $\sum_{ab} \hat \sigma_{ab}$ all double logarithms cancel. 

\section{LL resummation of the leading logarithms}
\subsection{Virtual corrections}\label{subsec:VirtLL}
To resum the electroweak Sudakov logarithms for for the virtual contributions we follow the work of~\cite{Chiu:2008vv,Chiu:2009mg}, which uses renormalization group equations in SCET. The calculation proceeds
in three steps. At the scale $\mu_Q^2 \sim s$, one matches the full theory onto 4-fermion
operators in SCET, where each of the fermion is represented by a different collinear sector in SCET. One then runs the theory from the scale $\mu_Q$ to the scale $\mu_V \sim m_V$, at which point the massive gauge bosons are integrated out of the theory and electroweak symmetry is broken.
As long as $s \gg m_V^2$, the mass of the vector boson can be set to zero in the matching onto SCET at the scale $\mu_Q$, and in the calculation of the anomalous dimension which governs the running from $\mu_Q$ to $\mu_V$. This implies that one can use an unbroken SU(2), simplifying these calculations substantially. 

In~\cite{Chiu:2008vv,Chiu:2009mg}, this resummation was carried out in full generality to NLL$^\prime$ accuracy. For the purposes of this work, we only work to LL accuracy, which simplifies the structure significantly. The first simplification is that only tree level matching is required both at the high and low scale. Furthermore, no operator mixing arises in the running from the high to the low scale. This will allow us to write relatively simple analytical formulae. 

As already mentioned, in the effective theory between $\mu_Q$ and $\mu_V$ one can use unbroken electroweak symmetry, such that there are a total of 7 operator that  contribute
\begin{align}
\label{eq:4fermionopsAbove}
{\cal L} = &  C_{QLT} \, Q^T L^T + C_{QLS} \, Q^S L^S +C_{ULS} \, U^S L^S 
+ C_{DLS} \, D^S L^S
\nonumber\\
& \qquad 
 + C_{QES} Q^S E^S
 + C_{UES} \, U^S E^S + C_{DES} \, D^S E^S 
\,,\end{align}
where we have defined the fermion bilinears in either triplet or singlet representation
\begin{align}
\label{eq:bilinears}
F^{S} = \bar F \gamma^\mu F\,,
\qquad 
F^{T} = \bar F \tau^a \gamma^\mu F
\,.
\end{align}
Here $Q$ and $L$ denote left-handed quark and lepton fields, respectively, while $U$, $D$ and $E$ denote the right handed up-type quarks, down-type 
quarks and electron fields. 
The tree level matching is given by
\begin{align}
s \, C^{(0)}_{QLT} & = 4 \pi \alpha_2
\nonumber\\
s \, C^{(0)}_{IFS} & = 4 \pi \alpha_1 \, Y_I Y_F
\,,\end{align}
where $I$ and $F$ denote the initial and final fermions of each singlet operator, and $Y_i$ is the hypercharge of particle $i$. The hypercharge normalization used is
\begin{align}
Y_i = Q_i - T^3_i
\label{eq:hypercharge}
\,,\end{align}
where $Q_i$ is the electromagnetic charge of the fermion $f=q / \ell$ and $T^3_i$ is the weak isospin.

In general, the renormalization group can mix these operators, however this mixing only arises starting at NLL. Since we are only interested in LL accuracy we can write
\begin{align}
\mu \frac{\df}{\df \mu} O_i(\mu) =\Gamma_i(\mu) \ln \frac{\mu^2}{s} O_i(\mu)
\,,
\end{align}
where the cusp anomalous dimension is
\begin{align}
\Gamma_i(\mu) = \sum_j\left[ \frac{\alpha_1(\mu)}{2 \pi} Y_j^2 + \frac{\alpha_2(\mu)}{2 \pi}T_j^2\right]_i
\,.
\end{align}
Here $[T_j]_i^2$ is the SU(2) Casimir of the j'th fermion (3/4 for left-handed fermions and 0 for right-handed fermions) in $O_i$. 

The solution to this RGE can easily be written down analytically, and one finds for the Wilson coefficients at the low scale
\begin{align}
C_i(\mu_V) = U^{LL}_i(\mu_V, \mu_Q) C_i(\mu_Q)
\,.
\end{align}
Working in the limit $\alpha_i \ln m_V^2 / s \sim 1$, one obtains the simple result
\begin{align}
U^{LL}_i(\mu_V, \mu_Q; s) = \frac{\exp\left[ - \Gamma_i(\sqrt{s}) \ln^2\frac{\mu_V}{\sqrt{s}}\right]}{\exp\left[ - \Gamma_i(\sqrt{s}) \ln^2\frac{\mu_Q}{\sqrt{s}}\right]}
\,.\end{align}
For $\mu_Q = \sqrt{s}$ and $\mu_V = m_V$ this result simplifies to
\begin{align}
U_i \equiv U^{LL}_i(m_V, \sqrt{s}; s) = \exp\left[ - \Gamma_i(\sqrt{s}) \ln^2 \frac{m_V}{\sqrt{s}}\right]
\,.\end{align}
Note that the anomalous dimension and therefore the evolution Kernel only depends on the type of initial and final state particles. Thus, we have 
\begin{align}
U_{IFS} = U_{IFT} \equiv U_{IF}
\,.
\end{align}

The SU(2) $ \otimes $ U(1) gauge structure is broken at $\mu_V=m_V$ to the U(1)$_{\rm em}$, and below that scale only the photon remains as gauge degree of freedom. Thus, below the scale $m_V$ one continues to run to a scale $\mu_\Lambda$, at which point the photon becomes unresolved. This running is determined by the cusp anomalous dimension
\begin{align}
\Gamma^{\rm em}(\mu) = \frac{\alpha_{\rm em}(\mu)}{2 \pi} Q_{\rm tot}^2
\,,
\end{align}
where 
\begin{align}
Q_{\rm tot}^2 \equiv \sum_i Q_i^2
\end{align}
is the sum of the square of the electromagnetic charges of all particles in the operator. The evolution Kernel between $\mu_V$ and $\mu_\Lambda$ is
\begin{align}
U^{\rm em}_{Q_{\rm tot}^2}\left(\mu_\Lambda, \mu_V; s \right)= \frac{\exp \left[ - \frac{\alpha_{\rm em}}{2\pi} Q_{\rm tot}^2 \ln^2 \frac{\mu_\Lambda}{\sqrt{s}} \right]}{\exp \left[ - \frac{\alpha_{\rm em}}{2\pi} Q_{\rm tot}^2 \ln^2 \frac{\mu_V}{\sqrt{s}} \right]}
\,.
\end{align}

We obtain the simple results for the resummed virtual corrections,
\begin{align}
\label{eq:virtualResummed}
\hat \sigma^{\rm LL}_{u\bar u \to e^-e^+} &= N \,  \frac{4\left(4\,U_{UL}^2+U_{QE}^2+16\,U_{UE}^2\right) \alpha_1^2 + U_{QL}^2 \left(\alpha_1+ 3 \alpha_2\right)^2}{54} \, \left[ U^{\rm em}_{26/9}\left(\Lambda, m_V; s \right) \right]^2 
\nn
\hat \sigma^{\rm LL}_{u\bar u \to \nu\bar\nu} &= N \,  \frac{16 \, U_{UL}^2 \alpha_1^2 + U_{QL}^2 \left(\alpha_1- 3 \alpha_2\right)^2}{54} \, \left[ U^{\rm em}_{8/9} \left(\Lambda, m_V; s \right) \right]^2 
\nn
\hat \sigma^{\rm LL}_{d\bar d \to e^-e^+} &= N \,  \frac{4 \left(U_{DL}^2 + U_{QE}^2 + 4 \, U_{DE}^2\right) \alpha_1^2 + U_{QL}^2 \left(\alpha_1- 3 \alpha_2\right)^2}{54} \, \left[ U^{\rm em}_{20/9}\left(\Lambda, m_V; s \right) \right]^2 
\nn
\hat \sigma^{\rm LL}_{d\bar d \to \nu \bar\nu} &= N \,  \frac{4 \, U_{DL}^2  \alpha_1^2 + U_{QL}^2 \left(\alpha_1+ 3 \alpha_2\right)^2}{54} \, \left[ U^{\rm em}_{2/9} \left(\Lambda, m_V; s \right) \right]^2
\nn
\hat \sigma^{\rm LL}_{u\bar d \to \nu e^+} &= N \,  \frac{2 \, U_{QL}^2 \alpha_2^2}{3} \, \left[ U^{\rm em}_{14/9}\left(\Lambda, m_V; s \right) \right]^2
\nn
\hat \sigma^{\rm LL}_{d\bar u \to e^-\nu} &= N \,  \frac{2 \, U_{QL}^2 \alpha_2^2}{3} \, \left[ U^{\rm em}_{14/9}\left(\Lambda, m_V; s \right) \right]^2
\,.\end{align}
Note that since $\alpha_1 / \alpha_2 = \tan^2(\theta_W) \sim 0.32$, the term proportional to $\alpha_1^2$ (which depends on  various different evolution Kernels) is numerically suppressed compared to the term proportional to $(\alpha_1 \pm 3 \alpha_2)^2$. Thus, to a good approximation, each leptonic final state gets the same suppression factor $U_{QL}$ from the resummation. 

A simple check of our results is that they reproduce the Born results given in~\eq{Born} if we set all resummation Kernels to unity, and that they reproduce the fixed order results in Eqs.(\ref{eq:FOVirtW}), (\ref{eq:FOVirtZ}), and (\ref{eq:FOVirtPhoton}) if we use the expansion $U_i^2 = 1 - 2\Gamma_i \ln^2 \frac{m_V}{\sqrt{s}} + \ldots$ and $\left[ U^{\rm em}_{Q_{\rm tot}^2}\left(\Lambda, m_V; s \right) \right]^2 = 1 + \frac{\alpha_{\rm em}(\mu)}{\pi} Q_{tot}^2 \left( \ln^2 \frac{m_V}{\sqrt{s}} - \ln^2 \frac{\Lambda}{\sqrt{s}} \right) + \ldots$. 

\subsection{Real corrections}\label{subsec:RealInclLL}

In this section we will calculate the resummation of the real emissions. We first give the results for the case of a single SU(2) gauge symmetry, and then extend the results to the case of the broken SU(2) $\otimes$ U(1) of the standard model.
\subsubsection{Simple SU(2)} \label{subsec:Simple SU(2)}
For a single SU(2) symmetry, the virtual results can be obtained from \eq{virtualResummed} by setting $\alpha_1 = \alpha_{\rm em} = 0$.  
It will be useful to rewrite them in a slightly different form, separating the contributions from the different helicities
\begin{align}
\label{eq:virtualAsShower}
\hat \sigma^{\rm LL}_{q_1^H q_2^H \to \ell_1^H \ell_2^H} &= \hat \sigma^{B}_{q_1^H q_2^H \to \ell_1^H \ell_2^H} 
\Delta^{\rm SU(2)}_{q_1^Hq_2^H\ell_1^H\ell_2^H}(m_V^2, s; s)
\,.\end{align}
The resummed logarithms are now contained in the factor $\Delta^{\rm SU(2)}_{q_1^Hq_2^H\ell_1^H\ell_2^H}(m_V^2, s; s)$.
The superscript $H$ denotes that each fermion has a fixed helicity. 
The Born cross-sections are given by
\begin{align}
\hat \sigma^{B}_{q^H q^H \to \ell^H \ell^H} &= N \,  \frac{8\, \alpha_2^2 \left( \, T^3_{q^H} T^3_{\ell^H}\right)^2}{3}
\nn
\hat \sigma^{B}_{q_1^L q_2^L \to \ell_1^L \ell_2^L} &= N \,  \frac{2\alpha_2^2 }{3}
\,,\end{align}
where $T^3_{q^H}$ denotes the weak isospin of the fermion $q / \ell$ with helicity $H$. The factor $\Delta^{\rm SU(2)}$resums the leading logarithms and is given by
\begin{align}
\label{eq:DeltaSU2}
\Delta^{\rm SU(2)}_{q_1^Hq_2^H\ell_1^H\ell_2^H}(m_V^2, s; s) = \exp\left[ - \frac{A^{\rm SU(2)}_{q_1^H q_1^H\ell_1^H\ell_2^H}}{2}\ln^2 \frac{m_V^2}{s}\right]
\,,\end{align}
where 
\begin{align}
\label{eq:AfHdef}
A^{\rm SU(2)}_{q_1^H q_1^H\ell_1^H\ell_2^H} = \frac{\alpha_2}{2\pi} \sum_iT_{i}^2
\,,\end{align}
and the sum over $i$ runs over all particles $i \in \{q_1^H, q_1^H, \ell_1^H, \ell_2^H$\}. 
Summing \eq{virtualAsShower} over all possible helicity structures, we reproduce \eq{virtualResummed} in the limit $\alpha_1 = \alpha_{\rm em} = 0$. 

By rewriting our result as in \eq{virtualAsShower}, one notices that it can be interpreted as the exclusive cross-section for the scattering process $q_1^H q_2^H \to \ell_1^H \ell_2^H$, where $\Delta^{\rm SU(2)}_{q_1^Hq_2^H\ell_1^H\ell_2^H}(m^2, s; s)$ is a Sudakov factor describing the probability of not having an emission of electroweak gauge bosons between the scales $s$ and $m_V^2$ for a process with center of mass energy $s$. Since of course the emission of a massive gauge boson always gives rise to a scale above $m_V^2$, this exclusive cross-section is by definition equal to the virtual result. 

\eq{virtualAsShower} is precisely the result that a parton shower would predict for the exclusive cross-section\footnote{Note that our Sudakov factor for the initial state particles does not include the ratios of PDFs that usually arise in backward evolution. This ratio of PDFs only contributes to NLL.}, and one can use insight from parton shower evolution to derive the expressions for real gauge boson radiation. The real emission of a gauge boson is given in a parton shower by the product of Altarelli-Parisi splitting functions, which describe the emission with a given transverse momentum $k_T^2$, multiplied by a Sudakov factor, which gives the no-branching probability above the value of $k_T^2$ as explained in \cite{Sjostrand:2006za}. 
%
Thus, the total inclusive real radiation cross-section (the cross section with one or more extra gauge bosons in the final state) is given by
\begin{align}
\hat \sigma^{\rm LL}_{q_1^H q_2^H \to \ell_1^H \ell_2^H + nV} &= \hat \sigma^{B}_{q_1^H q_2^H \to \ell_1^H \ell_2^H} \int_{m_V^2}^{s} \! \df k_T^2 \, \frac{\df}{\df k_T^2} 
\Delta^{\rm SU(2)}_{q_1^Hq_2^H\ell_1^H\ell_2^H}(k_T^2, s; s)
\nonumber\\
&= 
\hat \sigma^{B}_{q_1^H q_2^H \to \ell_1^H \ell_2^H} \left[ 1 - \Delta^{\rm SU(2)}_{q_1^Hq_2^H\ell_1^H\ell_2^H} (m_V^2, s; s)\right]
\,.\end{align}
Such an inclusive cross-section makes sense only if the measurement is not breaking the SU(2) symmetry. This is because the inclusive cross section is defined at a scale $\mu \sim k_T$, while the SU(2) symmetry is only broken at the scale $\mu \sim m_V$. This implies that the flavor structure one would obtain at the scale $k_T$ can be changed by the further emissions of extra gauge bosons, making an inclusive measurement with definite flavor structure (which is what breaks the symmetry) ill defined.  

Continuing to work in an unbroken SU(2) theory, one can also define the exclusive real radiation cross section (the cross section with exactly one extra gauge boson in the final state). This requires adding an extra no-branching probability from the scale $k_T^2$ to the scale $m_V^2$, which accounts for the fact that no extra gauge bosons are emitted from the fermions and the extra gauge boson with lower $k_T^2$. This extra factor is given by
\begin{align}
\Delta^{\rm SU(2)}_{q_1^Hq_2^H\ell_1^H\ell_2^HV}(m_V^2, k_T^2; s) \equiv \Delta_V(m_V^2, k_T^2; \hat k_T^2) \Delta^{\rm SU(2)}_{q_1^Hq_2^H\ell_1^H\ell_2^H}(m_V^2, k_T^2; s)
\,,
\end{align}
where the term $\Delta_V$ gives the probability of not emitting extra gauge bosons off the emitted vector boson
\begin{align}
\Delta_V(m_V^2, k_T^2; k_T^2) = \exp\left[ - \frac{\alpha_2 \, C_A }{4\pi}\ln^2 \frac{m_V^2}{k_T^2} \right] \,,
\end{align}
while the second term describes the no-emissions probability below $k_T$ off the fermions
\begin{align}
\Delta^{\rm SU(2)}_{q_1^Hq_2^H\ell_1^H\ell_2^H}(m_V^2, k_T^2; s) \equiv \frac{\Delta^{\rm SU(2)}_{q_1^Hq_2^H\ell_1^H\ell_2^H}(m_V^2, s; s)}{\Delta^{\rm SU(2)}_{q_1^Hq_2^H\ell_1^H\ell_2^H}(k_T^2,  s; s)}
\,.
\end{align}
Combining everything together, one therefore finds
\begin{align}
& \hat \sigma^{\rm LL}_{q_1^H q_2^H \to \ell_1^H \ell_2^H + V} 
\nn
&\qquad =  \hat \sigma^{B}_{q_1^H q_2^H \to \ell_1^H \ell_2^H}
 \int_{m_V^2}^{s} \! \df k_T^2 \, \frac{\df}{\df k_T^2} \left[\Delta^{\rm SU(2)}_{q_1^Hq_2^H\ell_1^H\ell_2^H}(k_T^2, s; s)\right] \Delta^{\rm SU(2)}_{q_1^Hq_2^H\ell_1^H\ell_2^HV}(m_V^2, k_T^2; s)
\nonumber\\
&\qquad =  \hat \sigma^{B}_{q_1^H q_2^H \to \ell_1^H \ell_2^H}
\, A^{\rm SU(2)}_{q_1^H q_1^H\ell_1^H\ell_2^H}\, \Delta^{\rm SU(2)}_{q_1^Hq_2^H\ell_1^H\ell_2^H}(m_V^2, s; s)  
\int_{m_V^2}^{s} \!\!\! \frac{\df k_T^2}{k_T^2} \ln \frac{s}{k_T^2} \Delta_{V}(m_V^2, k_T^2; k_T^2)
\,.\end{align}
The integral can be performed easily, and we write a general result 
\begin{align}
\label{eq:Idef}
I_\beta(m_V^2, s) &\equiv \int_{m_V^2}^{s} \! \frac{\df k_T^2}{k_T^2} \ln \frac{s}{k_T^2} \,\left[\Delta_{V}(m_V^2, k_T^2; k_T^2) \right]^\beta
\\
&= \frac{2\pi}{\alpha_2 \, \beta \, C_A} \left[ \frac{\sqrt{\alpha_2 \, \beta \, C_A}}{2} \ln\frac{m_V^2}{s} \, {\rm Erf}\left( \sqrt{\frac{\alpha_2 \, \beta \, C_A}{4\pi}} \, \ln\frac{m_V^2}{s}\right)   + \left[\Delta_V(m_V^2, s; s)\right]^\beta - 1 
\right] 
\,.
\nonumber
\end{align}
With this result, the exclusive cross section for a single emission is given by
\begin{align}
\hat \sigma^{\rm LL}_{q_1^H q_2^H \to \ell_1^H \ell_2^H + V} 
&=  \hat \sigma^{B}_{q_1^H q_2^H \to \ell_1^H \ell_2^H}
\, A^{\rm SU(2)}_{q_1^H q_1^H\ell_1^H\ell_2^H}\, \Delta^{\rm SU(2)}_{q_1^Hq_2^H\ell_1^H\ell_2^H}(m_V^2, s; s)  \, I_1(m_V^2,s)
 \,.
 \label{eq:Simple SU(2) cross section}
\end{align}

\subsubsection{Full SU(2) $\otimes$ U(1)}
We now extend the results of \subsec{Simple SU(2)} to include the full SU(2) $\otimes$ U(1) gauge structure. The Born cross-section is now given by
\begin{align}
\hat \sigma^{B}_{q^H q^H \to \ell^H \ell^H} &= N \,  \frac{8\, \left( \, \alpha_2 T^3_{q^H} T^3_{\ell^H} + \alpha_1 Y_{q^H} Y_{\ell^H} \right)^2}{3}
\nn
\hat \sigma^{B}_{q_1^L q_2^L \to \ell_1^L \ell_2^L} &= N \,  \frac{2\alpha_2^2 }{3}
\,,\end{align}
where as before $T^3_{f^H}$ denotes the weak isospin of the fermion $f=q / \ell$ with helicity $H$ and $Y_{f^H}$ denotes the hypercharge of the fermion $f=q / \ell$ with helicity $H$. The hypercharge normalization used is given by \eq{hypercharge}.

As in \subsec{Simple SU(2)}, we can write the LL cross section as the Born cross section times a Sudakov factor. However, contrary to the case of a single SU(2) symmetry, in the broken SU(2)$\otimes$U(1) symmetry of the standard model, below the scale $\mu = m_V$ one needs to continue to evolve the operators with the electromagnetic running. This gives
\begin{align}
\hat \sigma^{\rm LL}_{q_1^H q_2^H \to \ell_1^H \ell_2^H} &= \hat \sigma^{B}_{q_1^H q_2^H \to \ell_1^H \ell_2^H} 
\Delta_{q_1^Hq_2^H\ell_1^H\ell_2^H}(m_V^2, s; s) \, \Delta^{\rm em}_{q_1^Hq_2^H\ell_1^H\ell_2^H}(\Lambda^2, m_V^2; s)
\,.\end{align}
The Sudakov factor from $s$ to $m_V^2$ factors into 
two pieces, one for the SU(2) symmetry and one for the U(1) 
\begin{align}
\Delta_{q_1^Hq_2^H\ell_1^H\ell_2^H}(m_V^2, s; s) &=\Delta_{q_1^Hq_2^H\ell_1^H\ell_2^H}^{\rm SU(2)}(m_V^2, s; s) \, \Delta_{q_1^Hq_2^H\ell_1^H\ell_2^H}^{\rm U(1)}(m_V^2, s; s)
\,.\end{align}
The SU(2) contribution was given in \eq{DeltaSU2}, while the term coming from the U(1) symmetry is given by 
\begin{align}
 \Delta_{q_1^Hq_2^H\ell_1^H\ell_2^H}^{\rm U(1)}(m_V^2, s; s)) &=  \exp\left[ - \frac{A_{q_1^Hq_2^H\ell_1^H\ell_2^H}^{U(1)}}{2}\ln^2 \frac{m_V^2}{s} \right]
\,,\end{align}
with
\begin{align}
\label{eq:AfHdefU1}
A^{\rm U(1)}_{q_1^H q_1^H\ell_1^H\ell_2^H} = \frac{\alpha_1}{2\pi} \sum_iY_{i}^2
\,.\end{align}
The running below $m_V$ is determined only by the total charge of the operator, and one finds
\begin{align}
\Delta^{\rm em}_{q_1^Hq_2^H\ell_1^H\ell_2^H}(\Lambda^2, m_V^2; s) =  \exp\left[ - \frac{\alpha_{\rm em} Q_{\rm tot}^2}{4\pi}\left(\ln^2\frac{\Lambda^2}{s} - \ln^2\frac{m_V^2}{s} \right)\right]
\,.\end{align}
Summing over all possible helicity structures, we reproduce the resummed results of \subsec{VirtLL}.

To obtain the resummation of the real radiation, we follow the steps of \subsec{Simple SU(2)}, taking into account the full SU(2)$\otimes$U(1) structure above $m_V^2$ and the running due to the photon below $m_V^2$. For the $W^\pm$ bosons, the U(1) symmetry does not contribute, but one needs to be careful about the flavor structure when breaking the electroweak symmetry. 
One finds
\begin{align}
&\hat \sigma^{\rm LL}_{q_1^H q_2^H \to \ell_1^H \ell_2^H + W^\pm} 
\nn
&\quad
=  \left[ \Delta_{q_1^Hq_2^H\ell_1^H\ell_2^H}(m_V^2, s; s) \, \Delta^{\rm em}_{q_1^Hq_2^H\ell_1^H\ell_2^H W^\pm}(\Lambda^2, m_V^2; s) \int_{m_V^2}^{s} \!  \frac{\df k_T^2}{k_T^2} \ln \frac{s}{k_T^2} \Delta_{V}(m_V^2, k_T^2; k_T^2) \right]
\nonumber\\
&\quad\quad \times\left( \hat \sigma^{B}_{{q'}_1^H q_2^H \to \ell_1^H \ell_2^H} A^{W^\pm}_{q_1^H} + \hat \sigma^{B}_{q_1^H {q'}_2^H \to \ell_1^H \ell_2^H} A^{W^\pm}_{q_2^H} + \hat \sigma^{B}_{q_1^H q_2^H \to {\ell'}_1^H \ell_2^H} A^{W^\pm}_{\ell_1^H}+ \hat \sigma^{B}_{q_1^H q_2^H \to \ell_1^H {\ell'}_2^H} A^{W^\pm}_{\ell_2^H} \right) 
 \nonumber\\
&\quad
=  \left[ \Delta_{q_1^Hq_2^H\ell_1^H\ell_2^H}(m_V^2, s; s) \, \Delta^{\rm em}_{q_1^Hq_2^H\ell_1^H\ell_2^H W^\pm}(\Lambda^2, m_V^2; s) \, I_1 (m_V^2, s) \right]
\nonumber\\
&\quad\quad \times\left( \hat \sigma^{B}_{{q'}_1^H q_2^H \to \ell_1^H \ell_2^H} A^{W^\pm}_{q_1^H} + \hat \sigma^{B}_{q_1^H {q'}_2^H \to \ell_1^H \ell_2^H} A^{W^\pm}_{q_2^H} + \hat \sigma^{B}_{q_1^H q_2^H \to {\ell'}_1^H \ell_2^H} A^{W^\pm}_{\ell_1^H}+ \hat \sigma^{B}_{q_1^H q_2^H \to \ell_1^H {\ell'}_2^H} A^{W^\pm}_{\ell_2^H} \right) 
\,,
 \nonumber\\
 \label{eq:RealWFinal}
 \end{align}
where $f'$ is the fermion $f$ becomes after having radiated a $W^\pm$ that is $u'=d$, $d'=u$, $l'=\nu$ and $\nu'=l$ and for any flavor set which allows a $W^\pm$ emission there is one of the Born cross section which is zero because its electromagnetic charge is not conserved. Also, we broke $A^{\rm SU(2)}_{q_1^H q_1^H\ell_1^H\ell_2^H}$ in its component related to the emission of a $W^\pm$, that is
\begin{align}
A^{\rm SU(2)}_{q_1^H q_1^H\ell_1^H\ell_2^H} = A^{W^3}_{q_1^H q_1^H\ell_1^H\ell_2^H}+\sum_i A^{W^\pm}_i
 \,,
\end{align}
with
\begin{align}
A^{W^\pm}_{f^L}=\frac{\alpha_2}{4\pi}
\,,\qquad
A^{W^\pm}_{f^R}=0
 \,,
\qquad 
A^{W^3}_{q_1^H q_1^H\ell_1^H\ell_2^H}=\frac{\alpha_2}{2\pi} \sum_i (T_i^3)^2
 \,.
\end{align}
 

For the emissions of $Z$ bosons and photons, one needs to take into account the mixing between the third component of the SU(2) gauge symmetry and the U(1) gauge symmetry.
The emission of a $W^3$ boson is given by
\begin{align}
\hat \sigma^{\rm LL}_{q_1^H q_2^H \to \ell_1^H \ell_2^H + W^3} 
&=  \hat \sigma^{B}_{q_1^H q_2^H \to \ell_1^H \ell_2^H}
A^{\rm W^3}_{q_1^H q_1^H\ell_1^H\ell_2^H} \, \Delta_{q_1^Hq_2^H\ell_1^H\ell_2^H}(m_V^2, s; s)  
\nonumber\\
& \qquad  \times 
\Delta^{\rm em}_{q_1^Hq_2^H\ell_1^H\ell_2^H}(\Lambda^2, m_V^2; s) \, 
\int \! \frac{\df k_T^2}{k_T^2} \ln \frac{s}{k_T^2} \Delta_{V}(m_V^2, k_T^2; k_T^2) \, ,
\end{align}
while for the emission of a U(1) boson $B$
\begin{align}
\hat \sigma^{\rm LL}_{q_1^H q_2^H \to \ell_1^H \ell_2^H + B} 
&=  \hat \sigma^{B}_{q_1^H q_2^H \to \ell_1^H \ell_2^H}
A^{\rm U(1)}_{q_1^H q_1^H\ell_1^H\ell_2^H} \, \Delta_{q_1^Hq_2^H\ell_1^H\ell_2^H}(m_V^2, s; s)  
\nonumber\\
& \qquad  \times 
\Delta^{\rm em}_{q_1^Hq_2^H\ell_1^H\ell_2^H}(\Lambda^2, m_V^2; s) \, 
\int \! \frac{\df k_T^2}{k_T^2} \ln \frac{s}{k_T^2}
\,.\end{align}
To combine these two expressions into the emission of a $Z$ boson, we have to mix the amplitude of the emission of a $W^3$ with the amplitude of the emission of a $B$. Each individual emission is given by breaking the coefficient $A^{W^3}_{q_1^H q_1^H\ell_1^H\ell_2^H}$ and $A^{\rm U(1)}_{q_1^H q_1^H\ell_1^H\ell_2^H}$ in their components for an emission from one fermion, and the amplitude is given by the square root of this emission. We mix the amplitudes using that
\begin{align}
Z = s_W B - c_W W^3
 \,,
\end{align}
where $c_W = \cos(\theta_W)$ and  $s_W = \sin(\theta_W)$. This gives
\begin{align}
\hat \sigma^{\rm LL}_{q_1^H q_2^H \to \ell_1^H \ell_2^H + Z} 
&=  \hat \sigma^{B}_{q_1^H q_2^H \to \ell_1^H \ell_2^H} \, \Delta_{q_1^Hq_2^H\ell_1^H\ell_2^H}(m_V^2, s; s)  \, \Delta^{\rm em}_{q_1^Hq_2^H\ell_1^H\ell_2^H}(\Lambda^2, m_V^2; s)
\nonumber\\
& \qquad \times \int_{m_V^2}^{s} \! \frac{\df k_T^2}{k_T^2} \ln \frac{s}{k_T^2}
 R_{q_1^H q_2^H \ell_1^H \ell_2^H}(m_V^2, k_T^2)
 \,,
\end{align}
where we have defined
\begin{align}
R_{q_1^H q_2^H \ell_1^H \ell_2^H}(m_V^2, k_T^2) = \alpha_W \sum_i \left(s_W \sqrt{\frac{A^{U(1)}_{i}}{\alpha_W}} - c_W \sqrt{\frac{A^{W^3}_{i}}{\alpha_W} \Delta_{W}(m_V^2, k_T^2; k_T^2)} \right)^2
 \,,
\end{align}
with
\begin{align}
A^{\rm U(1)}_{f^H} = \frac{\alpha_1 }{2\pi}Y_{f^H}^2 \, , \qquad A^{W^3}_{f^H} = \frac{\alpha_1 }{2\pi}(T^3_{f^H})^2
 \,.
\end{align}
By using the simple relations
\begin{align}
\sqrt{\frac{A^{\rm U(1)}_{f^H}}{\alpha_W}} = \sqrt{\frac{1}{2\pi}} \frac{s_W}{c_W} Y_{f^H}
\,,
\qquad
\sqrt{\frac{A^{W^3}_{f^H}}{\alpha_W}} = \sqrt{\frac{1}{2\pi}} T^3_{f^H}
\,,\end{align}
we obtain the final result for the $Z$ boson emission cross section
\begin{align}
\hat \sigma^{\rm LL}_{q_1^H q_2^H \to \ell_1^H \ell_2^H + Z} 
&=  \hat \sigma^{B}_{q_1^H q_2^H \to \ell_1^H \ell_2^H} \, \Delta_{q_1^Hq_2^H\ell_1^H\ell_2^H}(m_V^2, s; s)  \, \Delta^{\rm em}_{q_1^Hq_2^H\ell_1^H\ell_2^H}(\Lambda^2, m_V^2; s)
\nn\\
& \qquad
\times \int_{m_V^2}^{s} \! \frac{\df k_T^2}{k_T^2} \ln \frac{s}{k_T^2}
\Bigg{(}
s_W^2 A^{\rm U(1)}_{q_1^Hq_2^H\ell_1^H\ell_2^H} - A^{\text{mixing}}_{q_1^Hq_2^H\ell_1^H\ell_2^H} \sqrt{\Delta_{W}(m_V^2, k_T^2; k_T^2)}
 \nonumber\\
& \qquad\qquad\qquad
 +c_W^2 A^{W^3}_{q_1^Hq_2^H\ell_1^H\ell_2^H} \Delta_{W}(m_V^2, k_T^2; k_T^2) \Bigg{)}  \nonumber\\
&= \hat \sigma^{B}_{q_1^H q_2^H \to \ell_1^H \ell_2^H} \, \Delta_{q_1^Hq_2^H\ell_1^H\ell_2^H}(m_V^2, s; s)  \, \Delta^{\rm em}_{q_1^Hq_2^H\ell_1^H\ell_2^H}(\Lambda^2, m_V^2; s)\, \nonumber\\ & \qquad\qquad
\Bigg{(} s_W^2 A^{\rm U(1)}_{q_1^Hq_2^H\ell_1^H\ell_2^H} \, \frac{1}{2} \ln^2 \frac{m_V^2}{s}
\nonumber\\
& \qquad\qquad
  - A^{\text{mixing}}_{q_1^Hq_2^H\ell_1^H\ell_2^H}
\, I_{\frac{1}{2}}(m_V^2,s)
+
c_W^2 \, A^{W^3}_{q_1^Hq_2^H\ell_1^H\ell_2^H} \, I_{1}(m_V^2,s)
\Bigg{)} \,, 
 \label{eq:RealZFinal}
\end{align}
where
\begin{align}
A^{\text{mixing}}_{q_1^Hq_2^H\ell_1^H\ell_2^H} = \frac{\alpha_{\rm em}}{\pi} \sum_{i}  T^3_{i} Y_{i}
\,.
\end{align}

For the emission of a photon, we use that, for a scale higher than the electroweak bosons masses, the photon is a mixing of the $B$ and $W^3$ bosons.
\begin{align}
\gamma = c_W B + s_W W^3
 \,,
\end{align}
while, for a scale lower than the electroweak bosons masses, the photon can still be produced proportionally to the derivative of its no-branching probability $\Delta^{\rm em}_{q_1^Hq_2^H\ell_1^H\ell_2^H}(k_T^2, m_V^2; s)$.
\begin{align}
\hat \sigma^{\rm LL}_{q_1^H q_2^H \to \ell_1^H \ell_2^H + \gamma} 
&= \hat \sigma^{B}_{q_1^H q_2^H \to \ell_1^H \ell_2^H} \, \Delta_{q_1^Hq_2^H\ell_1^H\ell_2^H}(m_V^2, s; s)
\nonumber\\
& \qquad \times \Bigg{[}
\Delta^{\rm em}_{q_1^Hq_2^H\ell_1^H\ell_2^H}(\Lambda^2, m_V^2; s)
\bigg{(}
c_W^2 A^{\rm U(1)}_{q_1^Hq_2^H\ell_1^H\ell_2^H} \frac{1}{2}\log^2(\frac{m_V^2}{s})
\nonumber\\
& \qquad  \qquad + A^{\text{mixing}}_{q_1^Hq_2^H\ell_1^H\ell_2^H}
I_{\frac{1}{2}}(m_V^2,s) +
s_W^2 A^{W^3}_{q_1^Hq_2^H\ell_1^H\ell_2^H} 
I_{1}(m_V^2,s)
\bigg{)} +
\nonumber\\
& \qquad \qquad \int_{\Lambda^2}^{m_V^2} \! \df k_T^2 \! \frac{\df}{\df k_T^2} \left[\Delta^{\rm em}_{q_1^Hq_2^H\ell_1^H\ell_2^H}(k_T^2, m_V^2; s) \right] \Delta^{\rm em}_{q_1^Hq_2^H\ell_1^H\ell_2^H}(\Lambda^2, k_T^2; s)
\Bigg{]} 
\nonumber\\
&= \hat \sigma^{B}_{q_1^H q_2^H \to \ell_1^H \ell_2^H} \, \Delta_{q_1^Hq_2^H\ell_1^H\ell_2^H}(m_V^2, s; s) \Delta^{\rm em}_{q_1^Hq_2^H\ell_1^H\ell_2^H}(\Lambda^2, m_V^2; s)
\nonumber\\
& \qquad \times \Bigg{[}
c_W^2 A^{\rm U(1)}_{q_1^Hq_2^H\ell_1^H\ell_2^H} \frac{1}{2}\log^2(\frac{m_V^2}{s})+ A^{\text{mixing}}_{q_1^Hq_2^H\ell_1^H\ell_2^H}
I_{\frac{1}{2}}(m_V^2,s) +
\nonumber\\
& \qquad  \qquad  s_W^2 A^{W^3}_{q_1^Hq_2^H\ell_1^H\ell_2^H} 
I_{1}(m_V^2,s) + \frac{\alpha Q^2_{\rm tot}}{4 \pi} \left( \ln^2 \frac{\Lambda^2}{s} - \ln^2 \frac{m_V^2}{s} \right)
\Bigg{]} 
\,.
\label{eq:RealPhotonFinal}
\end{align}

\section{Results}\label{sec:Results}
\begin{figure}[h!]
\centering
\includegraphics[width=0.9\textwidth]{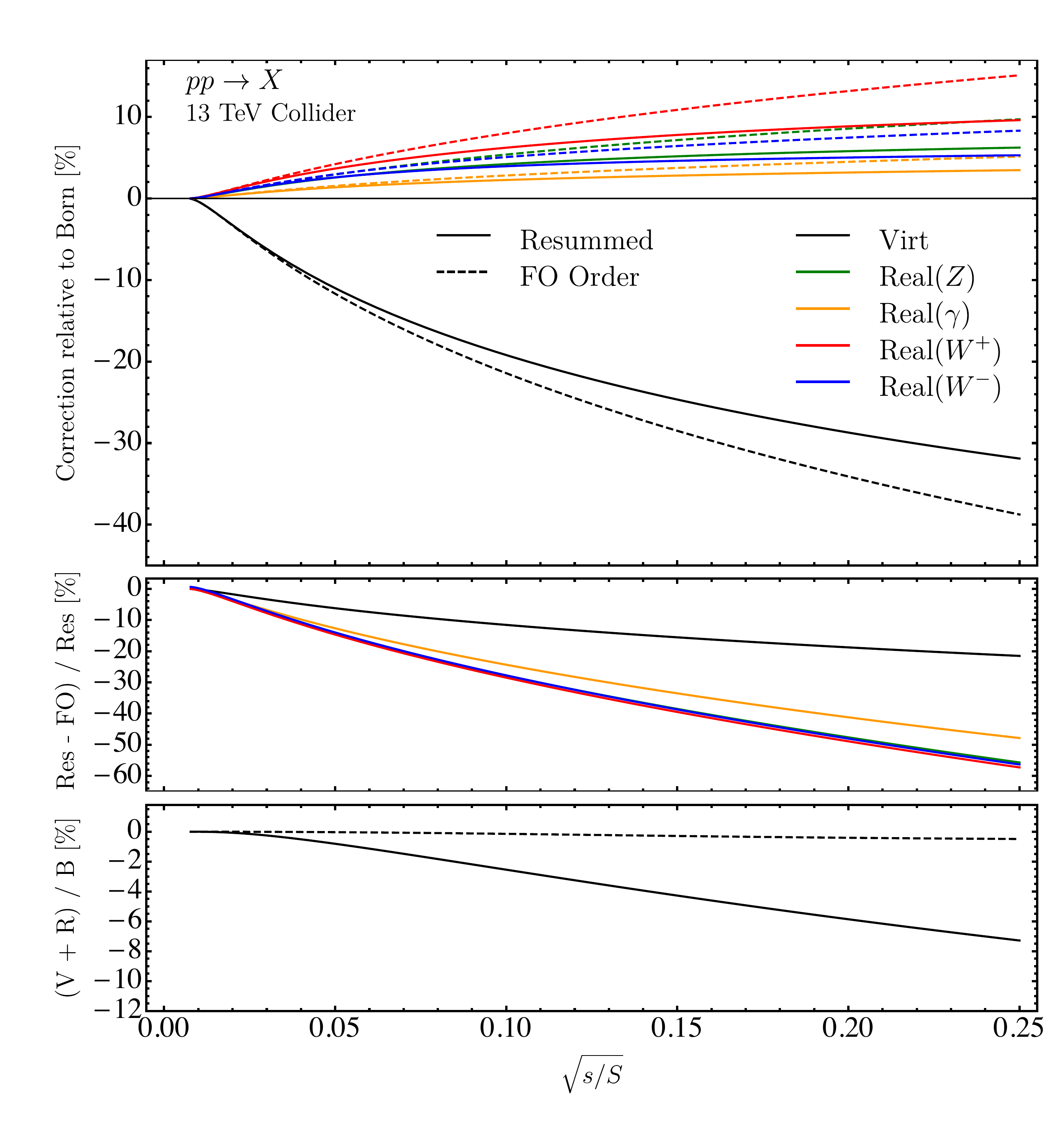}
\caption{The cross-section summed over all lepton flavors for the 13 TeV LHC. On the top we show the individual corrections relative to the Born cross-section as defined in~\eq{upperplot}, in the middle the relative size of the resummation as defined in \eq{middleplot}, while on the bottom we show the total perturbartive correction relative to the Born as defined in \eq{lowerplot}. 
Virtual corrections are shown in black, while real corrections with a $Z$, photon, $W^+$, $W^-$ are shown in green, orange, red and blue. Resummed corrections are shown in solid lines, while fixed order results are dashed. The x-axis denotes the fraction of the partonic center of mass energy relative to the collider center of mass energy. }
\label{fig:Inclusive13}
\end{figure}
\begin{figure}[h!]
\centering
\includegraphics[width=0.9\textwidth]{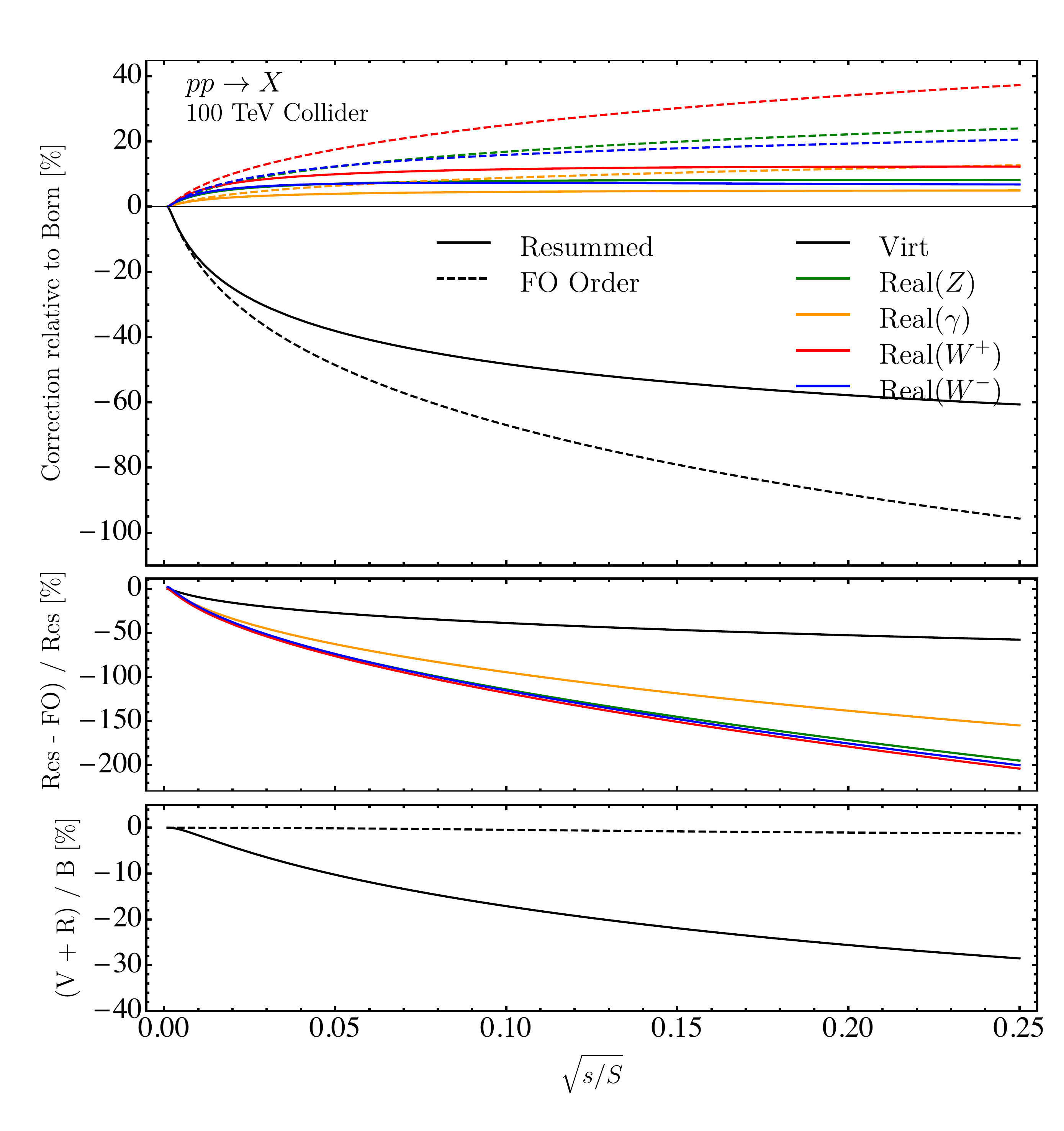}
\caption{The cross-section summed over all lepton flavors for a 100 TeV $pp$ collider on the right. All colors are the same as in \fig{Inclusive13}.}
\label{fig:Inclusive100}
\end{figure}
\begin{figure}[h!]
\centering
\subfloat[13 TeV LHC]{\includegraphics[width=0.5\textwidth]{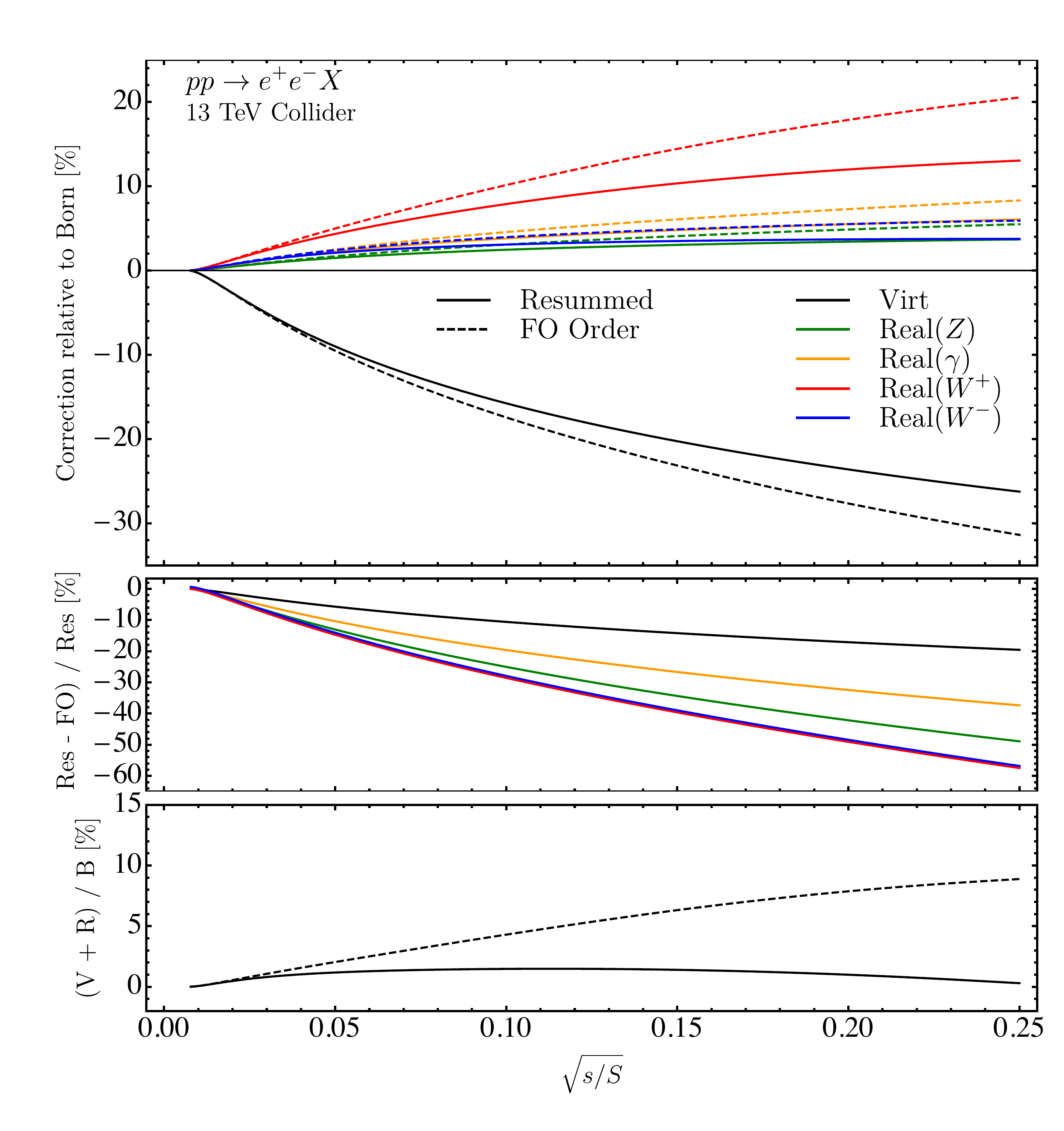}}
\hfill%
\subfloat[100 TeV collider]{\includegraphics[width=0.5\textwidth]{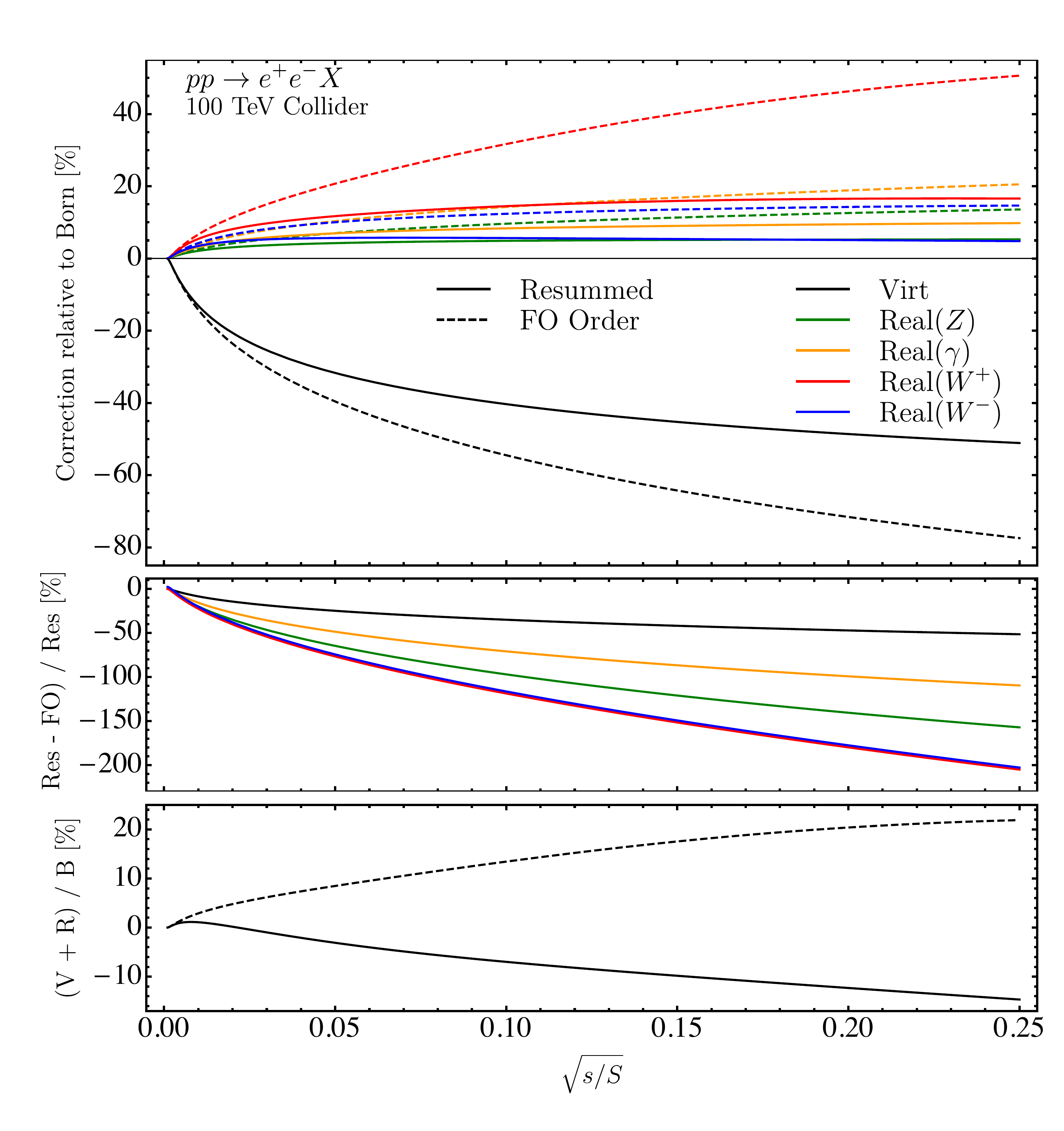}}
\caption{The cross-section for $e^+ e^-$ for the 13 TeV LHC on the left, and 100 TeV $pp$ collider on the right. 
All colors are the same as in \fig{Inclusive13}. Note that the scaling of the $y$ axis is different for the LHC and the 100 TeV collider.}
\label{fig:ee}
\end{figure}
\begin{figure}[h!]
\centering
\subfloat[13 TeV LHC]{\includegraphics[width=0.5\textwidth]{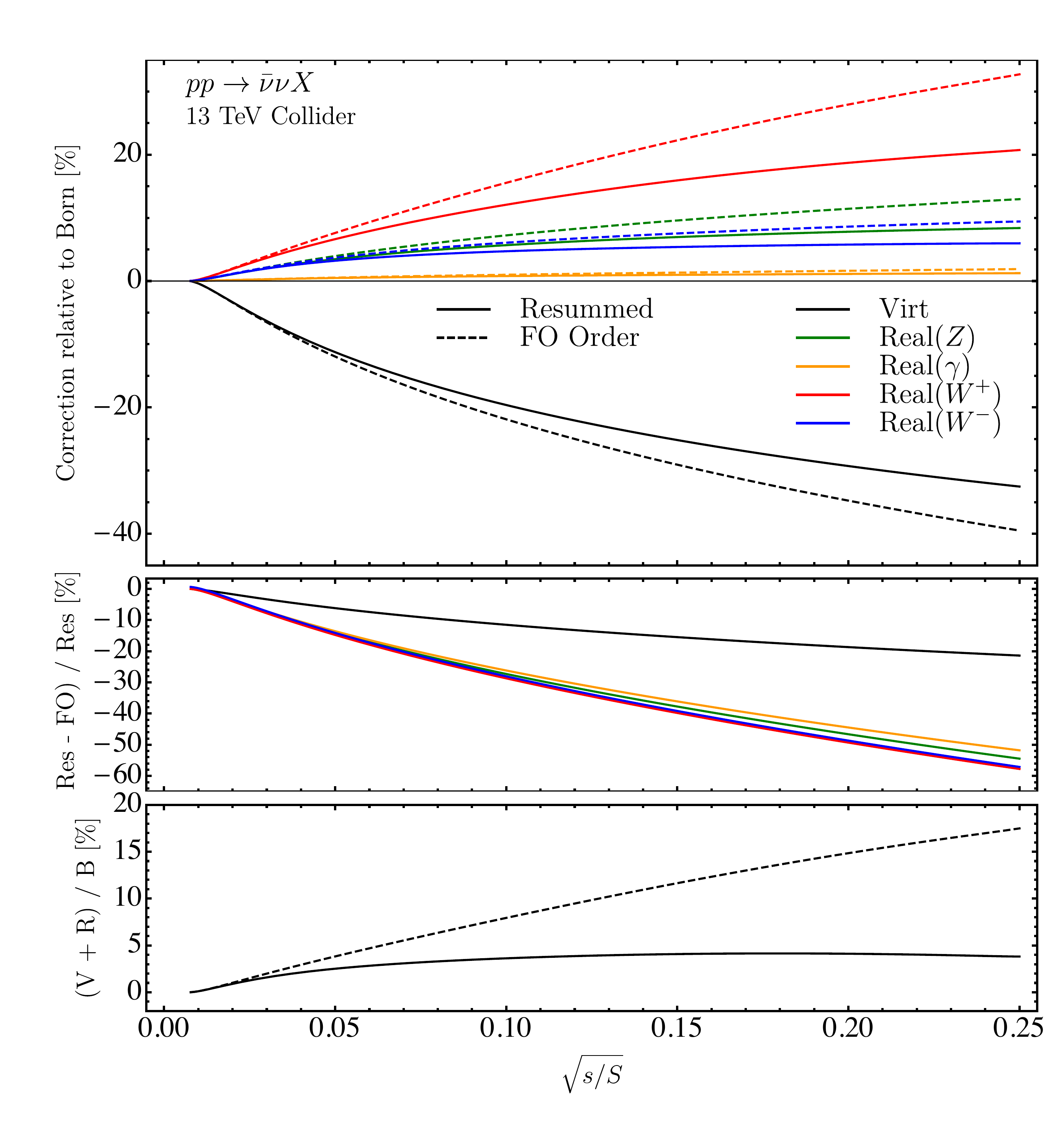}}
\hfill%
\subfloat[100 TeV collider]{\includegraphics[width=0.5\textwidth]{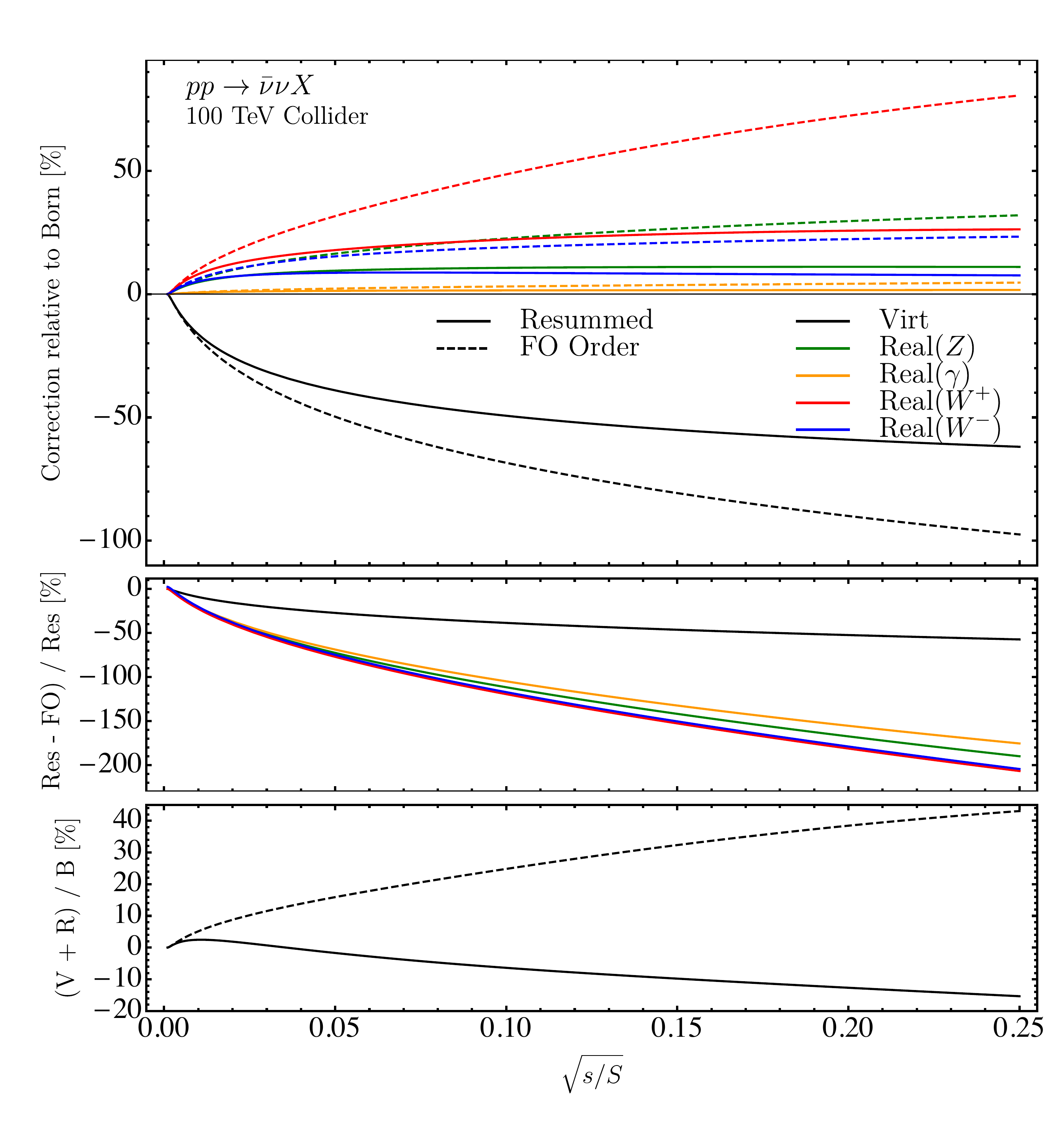}}
\caption{The cross-section for $\bar \nu\nu$ for the 13 TeV LHC on the left, and 100 TeV $pp$ collider on the right. All colors are the same as in \fig{Inclusive13}.}
\hspace{-2ex}
\label{fig:nunu}
\end{figure}
\begin{figure}[h!]
\centering
\subfloat[13 TeV LHC]{\includegraphics[width=0.5\textwidth]{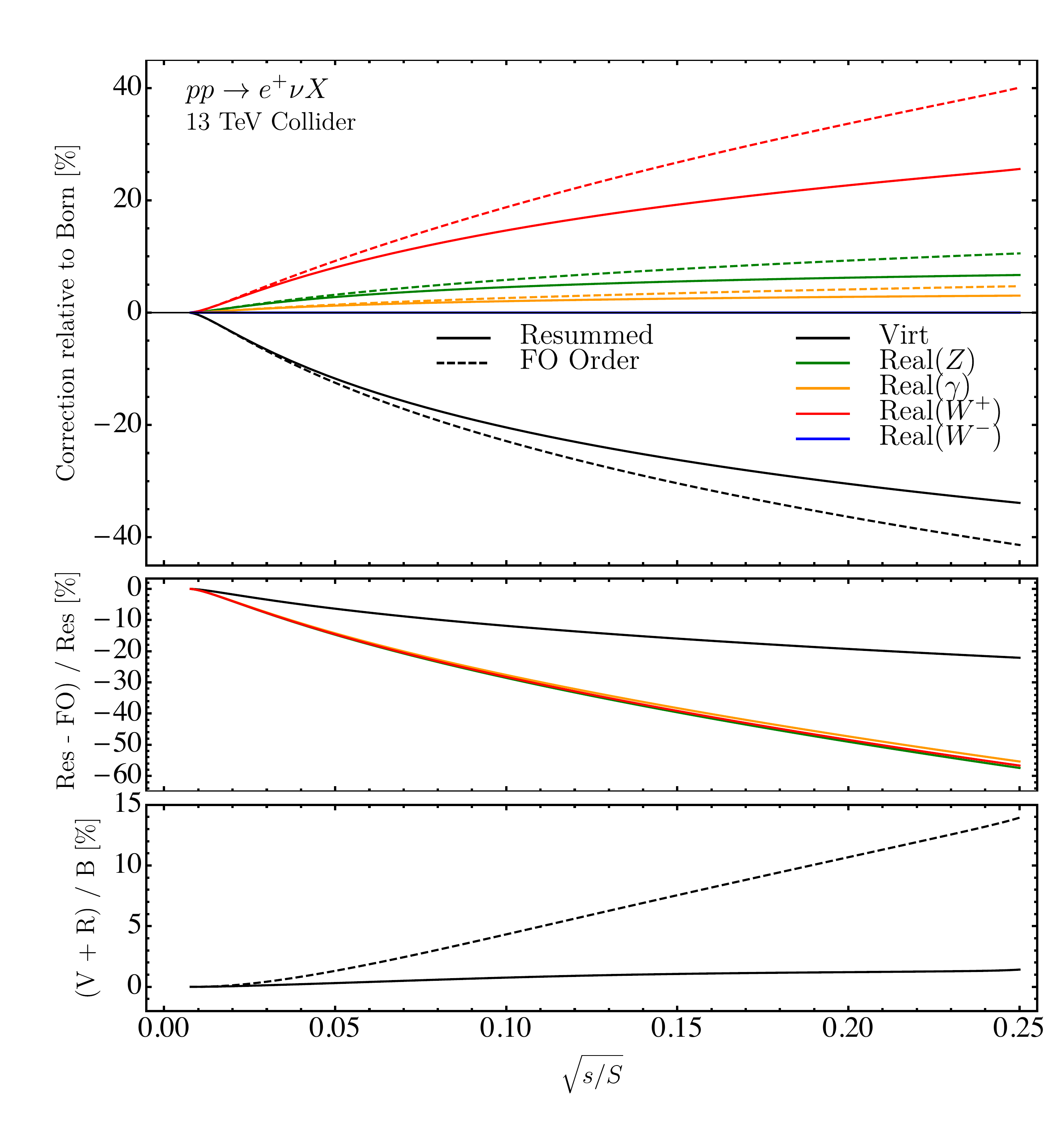}}
\hfill%
\subfloat[100 TeV collider]{\includegraphics[width=0.5\textwidth]{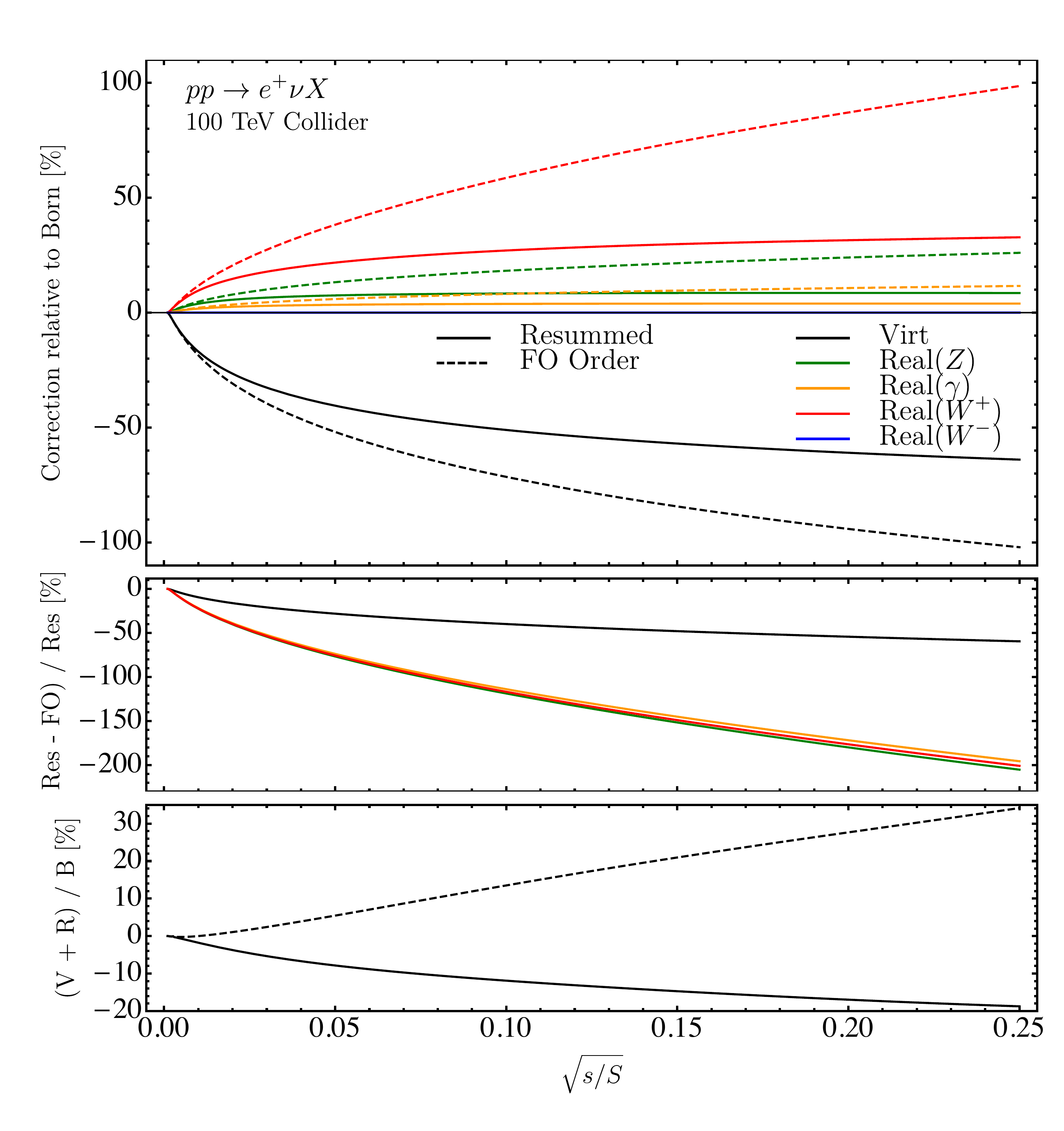}}
\caption{The cross-section summed for $e^+ \nu$ for the 13 TeV LHC on the left, and 100 TeV $pp$ collider on the right. All colors are the same as in \fig{Inclusive13}.}
\hspace{-2ex}
\label{fig:enu}
\end{figure}
\begin{figure}[h!]
\centering
\subfloat[13 TeV LHC]{\includegraphics[width=0.5\textwidth]{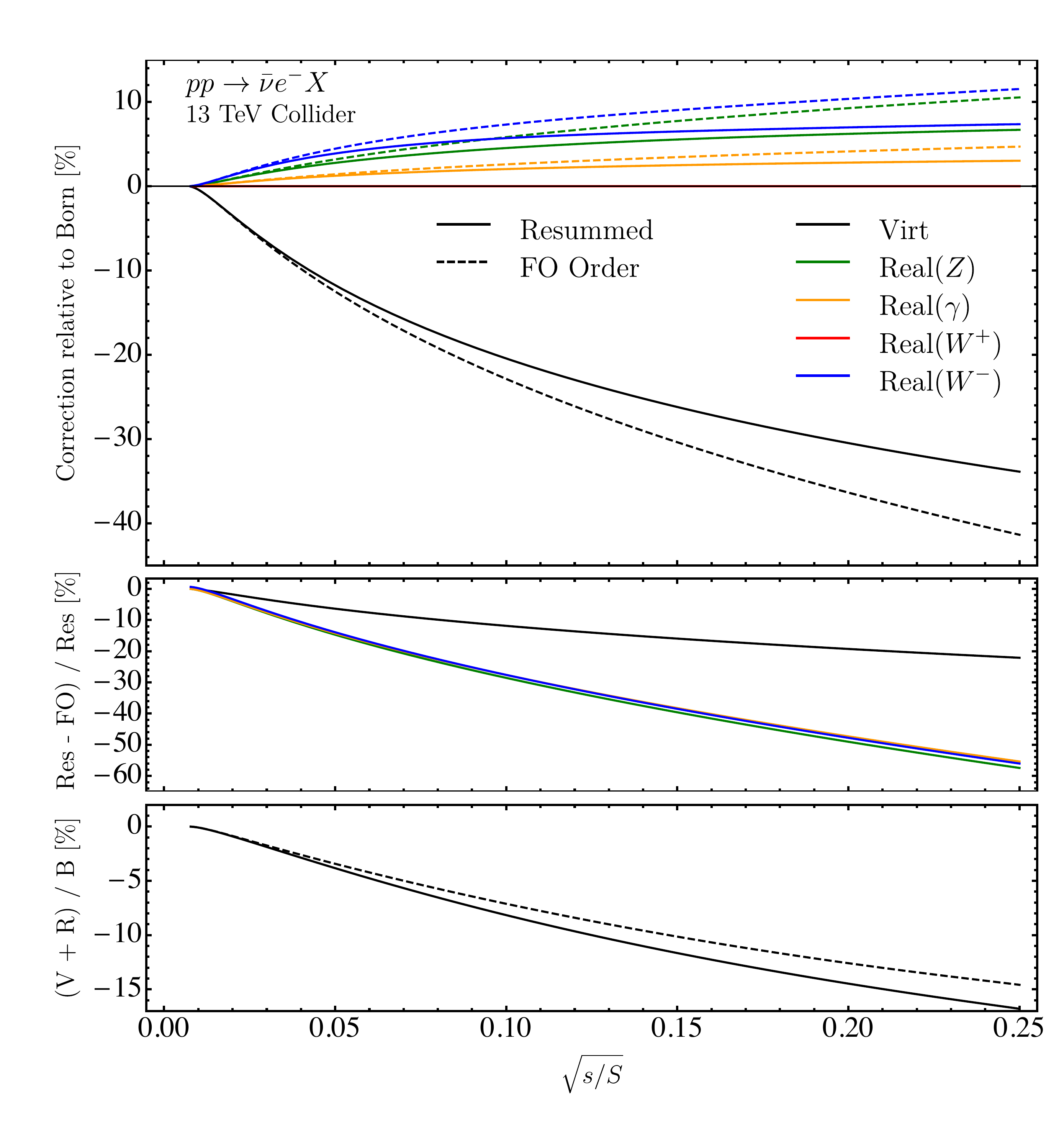}}
\hfill%
\subfloat[100 TeV collider]{\includegraphics[width=0.5\textwidth]{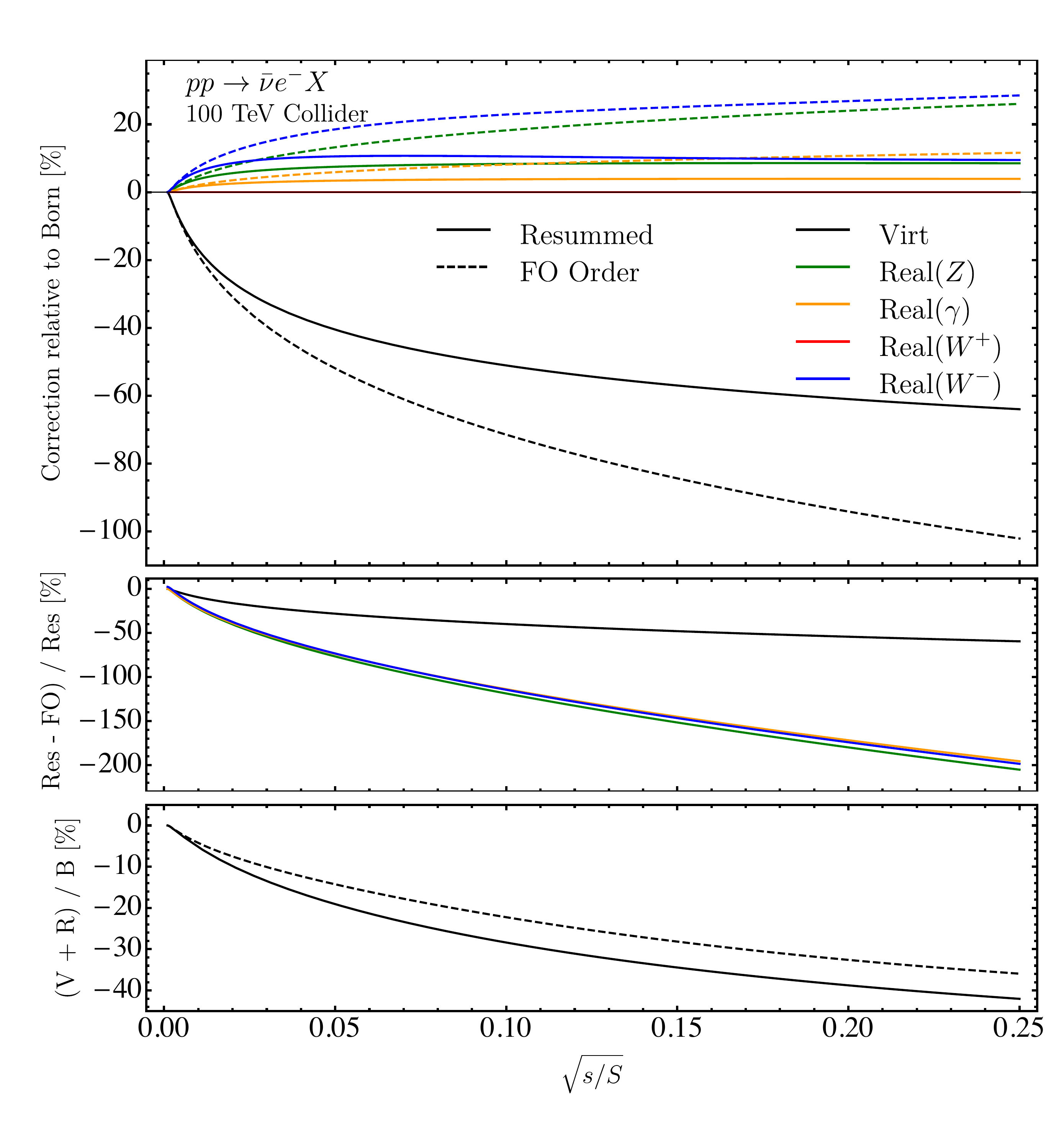}}
\caption{The cross-section for $\bar\nu e^-$ for the 13 TeV LHC on the left, and 100 TeV $pp$ collider on the right.  All colors are the same as in \fig{Inclusive13}.}
\hspace{-2ex}
\label{fig:nue}
\end{figure}

In this section we analyze the results presented in the last two sections
for both the the 13 TeV LHC and a 100 TeV proton-proton collider. The main reason to present the results for a 100 TeV collider is that it is the main candidate to succeed to the LHC and the importance of our results increase with the energy. Our results show that resumming both the real and virtual logarithms is essential for a 100 TeV collider.

We begin by explaining the format used in all of our plots: solid lines represent the resummed LL corrections, while dashed lines are the fixed order double logarithmic correction. Black lines correspond to the virtual corrections while the real corrections with $Z$, photon, $W^+$ and $W^-$ are shown in red, orange, blue and green, respectively. For each plot, we show on the top the perturbative corrections relative to the Born, in the middle the relative size of the resummation and on the bottom the total perturbative correction after summing virtual and real relative to the Born. To be more precise, on the top we plot for virtual and real
\begin{align}
\label{eq:upperplot}
{\rm Virt:} \quad \frac{\sigma_{pp \to \ell_1 \ell_2}-\sigma^{\rm B} _{pp \to \ell_1 \ell_2}}{\sigma^{\rm B} _{pp \to \ell_1 \ell_2}}
\, , \qquad\qquad
{\rm Real\, (V):} \quad \frac{\sigma_{pp \to \ell_1 \ell_2V}}{\sigma^{\rm B} _{pp \to \ell_1 \ell_2}}
\,,\end{align}
using either the resummed or fixed order expression.  In the middle we show for the virtual contribution
\begin{align}
\label{eq:middleplot}
{\rm Virt:} \quad \frac{\sigma^{\rm FO}_{pp \to \ell_1 \ell_2} - \sigma^{\rm Res}_{pp \to \ell_1 \ell_2}}{\sigma^{\rm Res}_{pp \to \ell_1 \ell_2}}
 \,, \qquad \qquad 
{\rm Real \, (V):}\quad 
\frac{\sigma^{\rm FO}_{pp \to \ell_1 \ell_2 V} - \sigma^{\rm Res}_{pp \to \ell_1 \ell_2 V}}{\sigma^{\rm Res}_{pp \to \ell_1 \ell_2 V}}
\,,\end{align}
while for the lower plot we show
\begin{align}
\label{eq:lowerplot}
\frac{\sigma_{pp \to \ell_1 \ell_2} + \sigma_{pp \to \ell_1 \ell_2V}-\sigma^{\rm B} _{pp \to \ell_1 \ell_2}}{\sigma^{\rm B} _{pp \to \ell_1 \ell_2}}
\,,\end{align}
for both fixed order and resummed. All effects are shown as a percentage.

We start with the result for the fully inclusive cross-section, where we sum over the flavors of all final state particles. In other words, we are summing over the virtual corrections for any lepton flavor, and the real corrections for any lepton flavor and gauge boson type.
In terms of equations, for the fixed order result, the virtual are obtained by summing over all terms in Eqs. (\ref{eq:FOVirtW}), (\ref{eq:FOVirtZ}), and  (\ref{eq:FOVirtPhoton}); while the real are given by Eqs. (\ref{eq:FORealZ}), (\ref{eq:RealWp}), (\ref{eq:RealWm}), and (\ref{eq:RealPhoton}). The resummed result are given by Eqs. (\ref{eq:virtualResummed}), (\ref{eq:RealZFinal}), (\ref{eq:RealWFinal}), and (\ref{eq:RealPhotonFinal}). All those results are for $\Lambda=m_Z$, that is we are resolving the photon only down to the mass of the $Z$. A lower $\Lambda$ would reduce all the exclusive cross section but the real cross section for photon as calculated in eq. (\ref{eq:RealPhotonFinal}) which will increase up to a value of $\Lambda$ low enough than 
$\frac{\alpha Q^2_{\rm tot}}{4 \pi} \left( \ln^2 \frac{\Lambda^2}{s} - \ln^2 \frac{m_V^2}{s} \right) = 1 - c_W^2 A^{\rm U(1)}_{q_1^Hq_2^H\ell_1^H\ell_2^H} \frac{1}{2}\log^2(\frac{m_V^2}{s})- A^{\text{mixing}}_{q_1^Hq_2^H\ell_1^H\ell_2^H} I_{\frac{1}{2}}(m_V^2,s) - s_W^2 A^{W^3}_{q_1^Hq_2^H\ell_1^H\ell_2^H} I_{1}(m_V^2,s)$,
and then decrease if $\Lambda$ continues to be lowered.

The results for the virtual and real contributions both at fixed and resummed order are shown in \fig{Inclusive13} for the LHC and in \fig{Inclusive100} for the 100 TeV collider. For the LHC, the corrections from virtual contributions range from ${\cal O}(15\%)$ at $\sqrt{s} \sim 1$ TeV to ${\cal O}(30\%)$ at $\sqrt{s} \sim 3$ TeV, while the real corrections for a given gauge boson are about a factor of 3 smaller individually. However, as can be seen from the ratio plot in the middle, the relative effect of the resummation is more than twice bigger for the real correction compared to the virtual correction. This clearly shows that the size of the resummation effect cannot be inferred from the size of the fixed order correction alone. The relative effect of the resummation for the virtual reaches from ${\cal O}(10\%)$ at $\sqrt{s} \sim 1$ TeV to ${\cal O}(20\%)$ at $\sqrt{s} \sim 3$ TeV while  the relative effect of the resummation for the real reaches from ${\cal O}(20\%)$ at $\sqrt{s} \sim 1$ TeV to ${\cal O}(50\%)$ at $\sqrt{s} \sim 3$ TeV.

In the lower part of the plot for the fixed order, one can see that after summing the virtual and real, the perturbative corrections largely cancel, but a small effect at the  ${\cal O}(1\%)$ level persists. For a fully inclusive cross section the logarithmically enhanced virtual and real corrections cancel against each other, up to the fact that the $pp$ initial state is not an iso-singlet. The large cancellation can be understood from \eq{InclusivePartonic}, which shows that switching the flavor of initial state anti-quark changes the sign of the partonic cross-section. Since pdf's for sea quarks are similar in magnitude, one expects $f_{\bar u / p} \sim f_{\bar d / p}$, explaining the cancellation. For the resummed result on the other hand, there is no cancellation. That is because even if the initial state is an SU(2) singlet, the cancellation would occur only for an inclusive result, that is summing the virtual to the real for any number of gauge boson. The remaining correction for the resummed result is thus mostly due to the production of more than one gauge boson with also an order 1\% correction due to the initial state not being an SU(2) singlet. This  remaining correction ranges from $2\%$ at $\sqrt{s} \sim 5$ TeV to $7\%$ at $\sqrt{s} \sim 25$ TeV.

For the 100 TeV collider, the results are qualitatively the same, but given the much larger reach in energy, the overall size of the effects are much larger. The virtual contributions range from ${\cal O}(30\%)$ at $\sqrt{s} \sim 5$ TeV to ${\cal O}(60\%)$ at $\sqrt{s} \sim 25$ TeV, with the real corrections again roughly a factor of 3 smaller. The relative size of the resummation, is also much larger, and at $\sqrt{s} \sim 25$ TeV changes the result by ${\cal O}(50\%)$ for the virtual and by ${\cal O}(200\%)$ for the real corrections. Thus, at such large energies, resummation has to be included to get a reliable estimate of the effects, not only for virtual corrections but also for the real emissions. Once the virtual and real corrections are added at fixed order, the total corrections again are very small, at the percent level; while for the resummed result, the total correction ranges from ${\cal O}(10\%)$ at $\sqrt{s} \sim 5$ TeV to ${\cal O}(30\%)$ at $\sqrt{s} \sim 25$ TeV.

Next, we consider the results for final states with specific leptons flavors. From the Figures \ref{fig:ee} to \ref{fig:nue}, one can see that at the LHC, the virtual corrections range from ${\cal O}(15\%)$ at $\sqrt{s} \sim 1$ TeV to ${\cal O}(30\%)$ at $\sqrt{s} \sim 3$ TeV, with the exact numbers depending  on the leptonic final state chosen, while at a 100 TeV collider they can exceed $50\%$ at $\sqrt{s} \sim 25$ TeV. Resummation at the LHC changes the virtual corrections by ${\cal O}(10\%)$ at  $\sqrt{s} \sim 1$ TeV to ${\cal O}(20\%)$ at  $\sqrt{s} \sim 3$ TeV, while at a 100 TeV collider the effect can become as large as $50\%$. Resummation at the LHC changes the real corrections by ${\cal O}(20\%)$ at  $\sqrt{s} \sim 1$ TeV to ${\cal O}(60\%)$ at  $\sqrt{s} \sim 3$ TeV, while at a 100 TeV collider the effect can become as large as $200\%$.
After summing over virtual and real corrections, the remaining perturbative corrections grow with energy are much larger than in the fully inclusive case. This is of course expected, since by specifying the leptonic final state, we are not considering an inclusive final state any longer. 

\section{Conclusions}\label{sec:Conclusions}

In this paper, we considered Drell-Yan production of leptons at a hadron collider, and presented result for the logarithmic resummation at LL accuracy for electroweak corrections to the total cross-section in simple analytic form.

Using these analytical results, we have analyzed the size of the corrections numerically for the 13 TeV LHC and a possible future 100 TeV $pp$ collider. Our results show than the real resummation is as important as the virtual resummation and its importance increases fast with the partonic center of mass energy: at the LHC, the resummation effect over the fixed order goes from around 10\% at 1 TeV to around 20\% at 3 TeV for the virtual with any leptonic final state and goes from around 20\% at 1 TeV to around 50\% at 3 TeV for all kinds of real emissions with any leptonic final state. The resummed virtual over the Born cross-section goes from around 10\% at 1 TeV to 30\% at 3 TeV, while each resummed reals are around the third of the resummed virtual. At a future 100 TeV $pp$ collider, the resummation effects over the fixed order can be much larger, and are around 50\% for the virtual and 200\% for the real at 25 TeV in most cases. The resummed virtual over the Born cross-section is around 80\% at 25 TeV.

For final states with fixed lepton flavors, summing the virtual and real corrections does not reduce the magnitude of the double logarithmic terms, as expected for a result that is not inclusive. After summing over all lepton flavors, a large cancellation is observed at fixed order, but a small logarithmic sensitivity remains, due to the fact that protons are not SU(2) singlets. On the other hand, for the resummed results, a large effect remains because we are including only the first emission for the real. The cancellation will happen only for an inclusive result, that is summing the virtual and the real for any number of emitted bosons.

In this paper, we have only considered the resummation of the real corrections to the total cross-section, and not given results that are differential in the emitted vector boson. That implies that we were only able to present results that were differential in $s$, and not $m_{\ell\ell}$, since these differ for real emissions. However, one can use the same approach used in this work, namely using the analogy to a parton shower, to obtain results that are fully differential. This will be the subject of a future paper.

\acknowledgments
{We would like the thank Aneesh Manohar for several stimulating discussions, and Michelangelo Mangano for helpful comments. This work was supported by by the Office of High Energy Physics of the U.S. Department of Energy under contract DE-AC02-05CH11231. 
}

\bibliographystyle{JHEP}
\bibliography{References}

\providecommand{\href}[2]{#2}\begingroup\raggedright\begin{thebibliography}{10}

\bibitem{Kuhn:1999nn}
J.~H. Kuhn, A.~A. Penin and V.~A. Smirnov, \emph{{Summing up subleading Sudakov
  logarithms}}, \href{http://dx.doi.org/10.1007/s100520000462}{\emph{Eur. Phys.
  J.} {\bf C17} (2000) 97--105},
  [\href{http://arxiv.org/abs/hep-ph/9912503}{{\tt hep-ph/9912503}}].

\bibitem{Fadin:1999bq}
V.~S. Fadin, L.~N. Lipatov, A.~D. Martin and M.~Melles, \emph{{Resummation of
  double logarithms in electroweak high-energy processes}},
  \href{http://dx.doi.org/10.1103/PhysRevD.61.094002}{\emph{Phys. Rev.} {\bf
  D61} (2000) 094002}, [\href{http://arxiv.org/abs/hep-ph/9910338}{{\tt
  hep-ph/9910338}}].

\bibitem{Ciafaloni:1999ub}
P.~Ciafaloni and D.~Comelli, \emph{{Electroweak Sudakov form-factors and
  nonfactorizable soft QED effects at NLC energies}},
  \href{http://dx.doi.org/10.1016/S0370-2693(00)00121-0}{\emph{Phys. Lett.}
  {\bf B476} (2000) 49--57}, [\href{http://arxiv.org/abs/hep-ph/9910278}{{\tt
  hep-ph/9910278}}].

\bibitem{Beccaria:2000jz}
M.~Beccaria, F.~M. Renard and C.~Verzegnassi, \emph{{Top quark production at
  future lepton colliders in the asymptotic regime}},
  \href{http://dx.doi.org/10.1103/PhysRevD.63.053013}{\emph{Phys. Rev.} {\bf
  D63} (2001) 053013}, [\href{http://arxiv.org/abs/hep-ph/0010205}{{\tt
  hep-ph/0010205}}].

\bibitem{Hori:2000tm}
M.~Hori, H.~Kawamura and J.~Kodaira, \emph{{Electroweak Sudakov at two loop
  level}}, \href{http://dx.doi.org/10.1016/S0370-2693(00)01027-3}{\emph{Phys.
  Lett.} {\bf B491} (2000) 275--279},
  [\href{http://arxiv.org/abs/hep-ph/0007329}{{\tt hep-ph/0007329}}].

\bibitem{Ciafaloni:2000df}
M.~Ciafaloni, P.~Ciafaloni and D.~Comelli, \emph{{Bloch-Nordsieck violating
  electroweak corrections to inclusive TeV scale hard processes}},
  \href{http://dx.doi.org/10.1103/PhysRevLett.84.4810}{\emph{Phys. Rev. Lett.}
  {\bf 84} (2000) 4810--4813}, [\href{http://arxiv.org/abs/hep-ph/0001142}{{\tt
  hep-ph/0001142}}].

\bibitem{Denner:2000jv}
A.~Denner and S.~Pozzorini, \emph{{One loop leading logarithms in electroweak
  radiative corrections. 1. Results}},
  \href{http://dx.doi.org/10.1007/s100520100551}{\emph{Eur. Phys. J.} {\bf C18}
  (2001) 461--480}, [\href{http://arxiv.org/abs/hep-ph/0010201}{{\tt
  hep-ph/0010201}}].

\bibitem{Denner:2001gw}
A.~Denner and S.~Pozzorini, \emph{{One loop leading logarithms in electroweak
  radiative corrections. 2. Factorization of collinear singularities}},
  \href{http://dx.doi.org/10.1007/s100520100721}{\emph{Eur. Phys. J.} {\bf C21}
  (2001) 63--79}, [\href{http://arxiv.org/abs/hep-ph/0104127}{{\tt
  hep-ph/0104127}}].

\bibitem{Melles:2001ye}
M.~Melles, \emph{{Electroweak radiative corrections in high-energy processes}},
  \href{http://dx.doi.org/10.1016/S0370-1573(02)00550-1}{\emph{Phys. Rept.}
  {\bf 375} (2003) 219--326}, [\href{http://arxiv.org/abs/hep-ph/0104232}{{\tt
  hep-ph/0104232}}].

\bibitem{Beenakker:2001kf}
W.~Beenakker and A.~Werthenbach, \emph{{Electroweak two loop Sudakov logarithms
  for on-shell fermions and bosons}},
  \href{http://dx.doi.org/10.1016/S0550-3213(02)00171-2}{\emph{Nucl. Phys.}
  {\bf B630} (2002) 3--54}, [\href{http://arxiv.org/abs/hep-ph/0112030}{{\tt
  hep-ph/0112030}}].

\bibitem{Denner:2003wi}
A.~Denner, M.~Melles and S.~Pozzorini, \emph{{Two loop electroweak angular
  dependent logarithms at high-energies}},
  \href{http://dx.doi.org/10.1016/S0550-3213(03)00307-9}{\emph{Nucl. Phys.}
  {\bf B662} (2003) 299--333}, [\href{http://arxiv.org/abs/hep-ph/0301241}{{\tt
  hep-ph/0301241}}].

\bibitem{Pozzorini:2004rm}
S.~Pozzorini, \emph{{Next to leading mass singularities in two loop electroweak
  singlet form-factors}},
  \href{http://dx.doi.org/10.1016/j.nuclphysb.2004.05.025}{\emph{Nucl. Phys.}
  {\bf B692} (2004) 135--174}, [\href{http://arxiv.org/abs/hep-ph/0401087}{{\tt
  hep-ph/0401087}}].

\bibitem{Feucht:2004rp}
B.~Feucht, J.~H. Kuhn, A.~A. Penin and V.~A. Smirnov, \emph{{Two loop Sudakov
  form-factor in a theory with mass gap}},
  \href{http://dx.doi.org/10.1103/PhysRevLett.93.101802}{\emph{Phys. Rev.
  Lett.} {\bf 93} (2004) 101802},
  [\href{http://arxiv.org/abs/hep-ph/0404082}{{\tt hep-ph/0404082}}].

\bibitem{Jantzen:2005xi}
B.~Jantzen, J.~H. Kuhn, A.~A. Penin and V.~A. Smirnov, \emph{{Two-loop
  electroweak logarithms}}, \href{http://dx.doi.org/10.1103/PhysRevD.74.019901,
  10.1103/PhysRevD.72.051301}{\emph{Phys. Rev.} {\bf D72} (2005) 051301},
  [\href{http://arxiv.org/abs/hep-ph/0504111}{{\tt hep-ph/0504111}}].

\bibitem{Jantzen:2005az}
B.~Jantzen, J.~H. Kuhn, A.~A. Penin and V.~A. Smirnov, \emph{{Two-loop
  electroweak logarithms in four-fermion processes at high energy}},
  \href{http://dx.doi.org/10.1016/j.nuclphysb.2005.10.010,
  10.1016/j.nuclphysb.2006.07.004}{\emph{Nucl. Phys.} {\bf B731} (2005)
  188--212}, [\href{http://arxiv.org/abs/hep-ph/0509157}{{\tt
  hep-ph/0509157}}].

\bibitem{Jantzen:2006jv}
B.~Jantzen and V.~A. Smirnov, \emph{{The Two-loop vector form-factor in the
  Sudakov limit}},
  \href{http://dx.doi.org/10.1140/epjc/s2006-02583-9}{\emph{Eur. Phys. J.} {\bf
  C47} (2006) 671--695}, [\href{http://arxiv.org/abs/hep-ph/0603133}{{\tt
  hep-ph/0603133}}].

\bibitem{Chiu:2007yn}
J.-y. Chiu, F.~Golf, R.~Kelley and A.~V. Manohar, \emph{{Electroweak Sudakov
  corrections using effective field theory}},
  \href{http://dx.doi.org/10.1103/PhysRevLett.100.021802}{\emph{Phys. Rev.
  Lett.} {\bf 100} (2008) 021802}, [\href{http://arxiv.org/abs/0709.2377}{{\tt
  0709.2377}}].

\bibitem{Chiu:2008vv}
J.-y. Chiu, R.~Kelley and A.~V. Manohar, \emph{{Electroweak Corrections using
  Effective Field Theory: Applications to the LHC}},
  \href{http://dx.doi.org/10.1103/PhysRevD.78.073006}{\emph{Phys. Rev.} {\bf
  D78} (2008) 073006}, [\href{http://arxiv.org/abs/0806.1240}{{\tt
  0806.1240}}].

\bibitem{Manohar:2012rs}
A.~V. Manohar and M.~Trott, \emph{{Electroweak Sudakov Corrections and the Top
  Quark Forward-Backward Asymmetry}},
  \href{http://dx.doi.org/10.1016/j.physletb.2012.04.013}{\emph{Phys. Lett.}
  {\bf B711} (2012) 313--316}, [\href{http://arxiv.org/abs/1201.3926}{{\tt
  1201.3926}}].

\bibitem{Bauer:2001ct}
C.~W. Bauer and I.~W. Stewart, \emph{{Invariant operators in collinear
  effective theory}},
  \href{http://dx.doi.org/10.1016/S0370-2693(01)00902-9}{\emph{Phys. Lett.}
  {\bf B516} (2001) 134--142}, [\href{http://arxiv.org/abs/hep-ph/0107001}{{\tt
  hep-ph/0107001}}].

\bibitem{Bauer:2000ew}
C.~W. Bauer, S.~Fleming and M.~E. Luke, \emph{{Summing Sudakov logarithms in B
  ---> X(s gamma) in effective field theory}},
  \href{http://dx.doi.org/10.1103/PhysRevD.63.014006}{\emph{Phys. Rev.} {\bf
  D63} (2000) 014006}, [\href{http://arxiv.org/abs/hep-ph/0005275}{{\tt
  hep-ph/0005275}}].

\bibitem{Bauer:2000yr}
C.~W. Bauer, S.~Fleming, D.~Pirjol and I.~W. Stewart, \emph{{An Effective field
  theory for collinear and soft gluons: Heavy to light decays}},
  \href{http://dx.doi.org/10.1103/PhysRevD.63.114020}{\emph{Phys. Rev.} {\bf
  D63} (2001) 114020}, [\href{http://arxiv.org/abs/hep-ph/0011336}{{\tt
  hep-ph/0011336}}].

\bibitem{Bauer:2001yt}
C.~W. Bauer, D.~Pirjol and I.~W. Stewart, \emph{{Soft collinear factorization
  in effective field theory}},
  \href{http://dx.doi.org/10.1103/PhysRevD.65.054022}{\emph{Phys. Rev.} {\bf
  D65} (2002) 054022}, [\href{http://arxiv.org/abs/hep-ph/0109045}{{\tt
  hep-ph/0109045}}].

\bibitem{Baur:2006sn}
U.~Baur, \emph{{Weak Boson Emission in Hadron Collider Processes}},
  \href{http://dx.doi.org/10.1103/PhysRevD.75.013005}{\emph{Phys. Rev.} {\bf
  D75} (2007) 013005}, [\href{http://arxiv.org/abs/hep-ph/0611241}{{\tt
  hep-ph/0611241}}].

\bibitem{Manohar:2014vxa}
A.~Manohar, B.~Shotwell, C.~Bauer and S.~Turczyk, \emph{{Non-cancellation of
  electroweak logarithms in high-energy scattering}},
  \href{http://dx.doi.org/10.1016/j.physletb.2014.11.050}{\emph{Phys. Lett.}
  {\bf B740} (2015) 179--187}, [\href{http://arxiv.org/abs/1409.1918}{{\tt
  1409.1918}}].

\bibitem{Bell:2010gi}
G.~Bell, J.~H. Kuhn and J.~Rittinger, \emph{{Electroweak Sudakov Logarithms and
  Real Gauge-Boson Radiation in the TeV Region}},
  \href{http://dx.doi.org/10.1140/epjc/s10052-010-1489-x}{\emph{Eur. Phys. J.}
  {\bf C70} (2010) 659--671}, [\href{http://arxiv.org/abs/1004.4117}{{\tt
  1004.4117}}].

\bibitem{Kinoshita:1962ur}
T.~Kinoshita, \emph{{Mass singularities of Feynman amplitudes}},
  \href{http://dx.doi.org/10.1063/1.1724268}{\emph{J. Math. Phys.} {\bf 3}
  (1962) 650--677}.

\bibitem{Chiu:2009mg}
J.-y. Chiu, A.~Fuhrer, R.~Kelley and A.~V. Manohar, \emph{{Factorization
  Structure of Gauge Theory Amplitudes and Application to Hard Scattering
  Processes at the LHC}},
  \href{http://dx.doi.org/10.1103/PhysRevD.80.094013}{\emph{Phys. Rev.} {\bf
  D80} (2009) 094013}, [\href{http://arxiv.org/abs/0909.0012}{{\tt
  0909.0012}}].

\bibitem{Sjostrand:2006za}
T.~Sjostrand, S.~Mrenna and P.~Z. Skands, \emph{{PYTHIA 6.4 Physics and
  Manual}}, \href{http://dx.doi.org/10.1088/1126-6708/2006/05/026}{\emph{JHEP}
  {\bf 05} (2006) 026}, [\href{http://arxiv.org/abs/hep-ph/0603175}{{\tt
  hep-ph/0603175}}].

\end{thebibliography}\endgroup

\end{document}